\documentclass[aps,twocolumn,amsmath,amssymb,preprintnumbers]{revtex4}
\usepackage{amsmath} \usepackage{amsfonts} \usepackage{amssymb}
\usepackage{bbm}
\usepackage{epsfig}
\usepackage{graphics}
\usepackage{graphicx}
\usepackage{titlesec}
\usepackage{mathtools}
\usepackage{environ}
\newcommand{\be}{\begin{equation}}
\newcommand{\ee}{\end{equation}}
\newcommand{\ba}{\begin{eqnarray}}
\newcommand{\ea}{\end{eqnarray}}
\newcommand{\nn}{\nonumber}
\newcommand{\kr}{\rangle}
\newcommand{\kl}{\langle}

\newcommand{\K}{{\cal K}}
\newcommand{\B}{{\cal B}}
\newcommand{\cL}{{\cal L}}
\newcommand{\tee}{(t+\epsilon)}
\newcommand{\teee}{(t+2\epsilon)}

\newcommand{\tr}{_{\tau\rho}}

\newcommand{\bl}{\big(}
\newcommand{\br}{\big)}
\newcommand{\ti}{t_{in}}

\newcommand{\cD}{{\cal D}}

\newcommand{\cN}{{\cal N}}
\newcommand{\T}{(t)}
\newcommand{\x}{(t,x)}
\newcommand{\vx}{(t,\vec x)}

\titleformat{\subsection}[block]{\normalfont\bfseries}{\thesubsection.}{1ex}{}
\titlespacing{\subsection}{0pt}{10pt}{1pt}[0pt]
\titleformat*{\section}{\large\bfseries}
\renewcommand{\thesubsection}{\arabic{subsection}}

\begin{document}

\title[ ]{Fermions as generalized Ising models}

\author{C. Wetterich}
\affiliation{Institut  f\"ur Theoretische Physik\\
Universit\"at Heidelberg\\
Philosophenweg 16, D-69120 Heidelberg}

\begin{abstract}
We establish a general map between Grassmann functionals for fermions and probability or weight distributions for Ising spins. The equivalence between the two formulations is based on identical transfer matrices and expectation values of products of observables. The map preserves locality properties and can be realized for arbitrary dimensions. We present a simple example where a quantum field theory for free massless Dirac fermions in two-dimensional Minkowski space is represented by an asymmetric Ising  model on a euclidean square lattice. 
\end{abstract}

\maketitle

\section{Introduction}
\label{Introduction}

Models for fermions and Ising spins \cite{IMIM,IMIS,IMOn} share the same type of observables. For fermions the observables are discrete occupation numbers $n_\alpha(t,\vec x)$ that take values zero or one. For $n_\alpha(t,\vec x)=1$ a fermion of type $\alpha$ is present at time $t$ at space position $\vec x$, while for $n_\alpha(t,\vec x)=0$ no such fermion is present. Other observables can be built as functions of occupation numbers. Ising spins $s_\alpha (t,\vec x)$ of type $\alpha$ at position $(t,\vec x)$ can take values $\pm 1$. The simple relation to occupation numbers reads $n_\alpha(t,\vec x)=\bl s_\alpha \vx +1\br/2$. We consider well defined regularized models by taking the positions $\vx$ as points on some lattice. A priori no metric is introduced such that $t$ can equally be interpreted as a time or space position. The notion of neighborhood is introduced by specifying the ``interactions'' of the model. We will typically consider simple next-neighbor or diagonal interactions for the generalized Ising models and simple ``hopping terms'' for the discretized fermion models. At this level the general properties of observables in models for fermions or Ising spins are the same.

Based on these simple observations a map between generalized Ising models and fermions has been proposed in ref. \cite{IMCWF}. It has been based on the notion of ``classical wave functions'' \cite{IMCWCW} both for fermions and Ising spins. In the present paper we establish the map between fermions and Ising spins directly on the fundamental level of ``functional integrals''. We construct the general map for an arbitrary dimension of space(time) and present concrete simple examples. If Ising spins are considered as ``discrete bosons'' this map establishes a general equivalence of fermions and discrete bosons. 

For the generalized Ising model the basic probabilistic concept is the probability distribution which associates to each configuration of Ising spins or occupation numbers $\{n\}$ a probability $p[n]=p\bl\{n\}\br$. We generalize this to arbitrary weight functions $w[n]$ which may not only take positive values. The functional integral is a sum over all configurations of occupation numbers. Fermionic models are formulated in terms of Grassmann variables $\psi_\alpha\vx$. The Grassmann weight function $\tilde w[\psi]$ is an element of the Grassmann algebra constructed from all $\psi_\alpha\vx$. The functional integral is a Grassmann integral over all $\psi_\alpha\vx$. 

We construct an invertible map between $w[n]$ and $\tilde w[\psi]$. Equivalence between the formulation in the ``occupation number basis'' in terms of Ising spins, and the ``fermion basis'' in terms of Grassmann variables, is established by the map of observables between the two formulations. Every observable that is a function of Ising spins finds its counterpart in an element of the Grassmann algebra. The map between observables and the (standard) rules for the computation of expectation values from functional integrals are such that all expectation values are the same both in the occupation number basis and the fermion basis. 

The proposed map is very general. It is not restricted to any number of dimensions for the positions $\vx$. It also preserves locality properties. Local interactions of Ising spins are mapped to local interactions for fermions. Fermion-boson maps for particular systems, as one-dimensional chains or topological excitations \cite{TS,Col,ChS,Kit1,Kit2}, have been fruitfully exploited in the past. The map proposed in the present paper differs from those earlier examples of fermion-boson equivalence. It is simply a very direct association between $n_\alpha\vx$ and $\psi_\alpha\vx$ that does  not involve any particular excitations. In contrast to the earlier maps based on particular excitations our map is a strictly local association between the basic degrees of freedom. Ising spins or occupation numbers can be associated to bits of information. We will call the map between the fermion basis and the occupation number basis the ``fermion-bit map''. 

In its most general form the fermion-bit map associates to each fermionic model, as specified by an action $S[\psi]$ and boundary conditions, a weight function $w[n]$ for occupation numbers. In general, this weight function could assume negative values for certain configurations $\{n\}$. Of  particular interest are fermionic models that lead to positive weight functions which obey for all configurations $\{n\}$ of Ising spins $w[n]\geq 0$. Such weight functions can be normalized in order to define a probability distribution $p[n]$. The latter can be evaluated by many standard methods of statistical  physics, including numerical Monte-Carlo simulations. This opens the possibility of a direct numerical evaluation of observables for fermionic models. 

In the present paper we discuss simple examples for the fermion-bit map. In particular, a two-dimensional model of free Dirac fermions is characterized by the standard weight factor for Minkowski-signature 
\be\label{IN0}
K[\psi]=e^{-i S_M[\psi]},
\ee
with Lorentz-invariant action $S_M$ written in a continuum notation as
\be\label{IN1}
S_M=\int_{t,x}i\bar\psi\gamma^\mu\partial_\mu\psi. 
\ee
The Grassmann weight function $\tilde w[\psi]$ multiplies $K[\psi]$ by boundary terms. The equivalent generalized Ising model involves diagonal interactions on a square lattice
\be\label{IN2}
K[n]=e^{-S_E[n]},
\ee
with 
\ba\label{IN3}
S_E&=&-\frac\beta2\sum_{t,x}\big\{s_1(t+\epsilon,x+\epsilon)s_1\x\nn\\
&&+s_2(t+\epsilon,x-\epsilon)s_2\x-2\big\}.
\ea
The two types of Ising spins $s_1,s_2$ correspond to the two components of the complex Grassmann variable $\psi\x=\bl \psi_1\x,\psi_2\x\br$. The interactions are asymmetric since $s_1$ and $s_2$ interact on diagonals in different directions. In the limit $\beta\to \infty$ the fermion--bit map relates $K[\psi]$ and $K[n]$ in eqs. \eqref{IN0} and \eqref{IN2}. This extends to boundary terms and observables. For suitable boundary terms the weight distribution $w[n]$ is positive. Free massless Dirac fermions in two dimensions can therefore be represented by the classical probability distribution of a generalized Ising model. In consequence, the fermionic observables of the model \eqref{IN0} can be computed by a Monte-Carlo simulation of the asymmetric Ising model \eqref{IN2}. 

This paper is organized as follows. In sect. \ref{Generalized Ising models} we recall basic properties of generalized Ising models in a formulation based on occupation numbers that is suitable for the fermion-bit map. We introduce local bilinear operators and the notion of a classical density matrix. This allows us to express all expectation values of observables that involve occupation numbers on one or two $t-$ layers in terms of the density matrix and operators, employing the standard quantum mechanical expression. In sect. \ref{Grassmann functional integral} we construct the equivalent Grassmann weight function and functional integral. We establish the equivalence of $w[n]$ and $\tilde w[\psi]$ by ensuring that the transfer matrix \cite{IMTM,IMMS,IMFU} is the same for the generalized Ising model and the associated fermion model. The map is completed in sect. \ref{Grassmann observables} by specifying the map for local bilinear operators between the occupation number basis and the fermion basis. This is extended to arbitrary products of local observables. 

In sect. \ref{Simple fermion models} we start with simple fermion models that are represented by generalized Ising models with a positive probability distribution $p[n]$. These ``unique jump chains'' can be associated with cellular automata \cite{IMCA,IC,IMTH,IMTH2,IMEL}. Sect. \ref{Two-dimensional fermions} shows that simple models of free massless fermions in two dimensions can be expressed by asymmetric Ising models with a positive probability distribution. We summarize our findings, discuss extensions to wider classes of fermion models, as well as a possible conceptual impact of our findings for the foundations of quantum mechanics and quantum field theory, in the concluding sect. \ref{Conclusions}.

\section{Generalized Ising models}
\label{Generalized Ising models}

Our basic variables are occupation numbers $n_\gamma(t)$ that can take values one or zero. Here $t$ is a discrete position variable taking $G$ values, $t_0\dots t_{G-1}$, that we assume equidistant, $t_0=\ti,~t_{n'}=\ti+n'\epsilon,~t_{G-1}=t_f$. The index $\gamma$ denotes different ``species'', $\gamma=1\dots M$. In a multi-dimensional setting it includes discrete position labels for the additional dimensions. Ising spins obey $s_\gamma=2n_\gamma-1$. A configuration $[n]$ is a set of $MG$ numbers one or zero. There are $2^{MG}$ different configurations.

\subsection{Weight function}

We define a weight function
\be\label{N1}
w[n]=\bar f_fK[n]f_{in}
\ee
as a function of the configurations $\{n\}$. We consider a quasi-local form
\be\label{N2}
K[n]=\prod^{t_f-\epsilon}_{t=\ti}\K(t),
\ee
with ``local factors'' $\K(t)$ depending on occupation numbers at $t$ and $t+\epsilon$, e.g. $n_\gamma(t)$ and $n_\gamma\tee$,
\be\label{2A}
\K(t)=\K\bl n\tee ,n(t)\br.
\ee
The product in eq. \eqref{N2} is over $G-1$ factors from $t_0=\ti$ to $t_{G-2}=t_f-\epsilon$. Boundary conditions are specified at $\ti$ by $f_{in}\bl n(\ti)\br$ depending on the occupation numbers $n_\gamma(\ti)$ at the ``initial boundary'', and similar at the ``final boundary'' by $\bar f_f\bl n(t_f)\br$. For positive $\K\T$ we write 
\be\label{3A}
\K\T=\exp \big\{-\cL\T\big\},
\ee
and 
\be\label{3B}
K[n]=\exp \big\{-S_E[n]\big\}=\exp \big\{-\sum_t\cL\T\big\}.
\ee
This form holds for generalized Ising models. We will, however, not restrict the discussion to positive $\K\T$.

The partition function $Z$ is given by a sum over all configurations
\be\label{N4}
Z=\int {\cal D}n w[n],
\ee
where we employ the language of a ``functional integral'' 
\be\label{14}
\int {\cal D}n=\sum_{\{n\}}=\prod_t\prod_\gamma\sum_{n_\gamma(t)=0,1}=\prod_t\int dn(t).
\ee
The product $\prod_t$ involves all $G$ values of $t$. If the weight function is positive semidefinite, $w[n]\geq 0$, a classical statistical probability distribution $p[n]$ is defined by
\be\label{N5}
p[n]=Z^{-1}w[n].
\ee
Observables $A[n]$ are real functions of the configurations $\{n\}$. Their expectation value is defined as
\be\label{N6}
\kl A \kr =Z^{-1}\int {\cal D}nw[n]A[n].
\ee
In particular, ``local observables'' $A(t)$ depend only on occupation numbers $n_\gamma(t)$ at a given $t$.

\subsection{Transfer matrix}

We will next express the weight function in terms of elements of the transfer matrix \cite{IMTM,IMMS,IMFU}. For this purpose we define a set of basis functions $f_\tau(t)$ that depend on occupation numbers $n_\gamma(t)$ at a given $t$, with $\tau=1\dots N,~N=2^M$. For a given set $[n(t)]$ of the $M$ ``local occupation numbers'' $n_\gamma(t)$ the basis functions $f_\tau$ involve a product of $M$ factors $n_\gamma$ or $(1-n_\gamma)$. If we label $\tau$ by an ordered sequence of numbers one or zero we define $f_\tau$ by inserting a factor $1-n_\gamma$ for each zero and a factor $n_\gamma$ for each one. For the example $\tau=(0,1,1,0,1,\dots)$ the basis function reads $f_\tau=(1-n_1)n_2n_3(1-n_4)n_5\dots$. The basis functions obey the product rule
\be\label{5}
f_\tau f_\rho=f_\tau\delta_{\tau\rho},
\ee
and the sum rule
\be\label{7}
\sum_\tau f_\tau=1.
\ee
For the integration over local configurations $[n(t)]$ one has
\be\label{6}
\int dnf_\tau=\prod_\gamma\sum_{n_\gamma=0,1} f_\tau=1,
\ee
implying
\be\label{N7}
\int dn f_\tau f_\rho=\delta_{\tau\rho}.
\ee

An arbitrary function $f(t)$ of the configurations of local occupation numbers $[n(t)]$ can be expanded in the basis functions $f_\tau(t)$,
\be\label{12}
f(t)=q_\tau(t)f_\tau(t).
\ee
Sums over repeated indices are implied if not stated explicitly otherwise. The transfer matrix $\bar S(t)$ is given by a double expansion of $\K (t)$ in $f_\tau\tee$ and $f_\rho(t)$
\be\label{N8}
\K(t)=f_\tau\tee \bar S\tr(t)f_\rho(t). 
\ee
By virtue of the product rule \eqref{5} we observe 
\be\label{N9}
\K\tee \K (t)=\sum_{\tau,\sigma,\rho}\quad
\bar S_{\tau\sigma}\tee \bar S_{\sigma\rho}(t)f_\tau\teee f_\sigma\tee f_\rho(t). 
\ee
This extends to arbitrary products of subsequent $\K(t')$ factors
\ba\label{N10}
&&\bar K(t_m, t)=
\prod^{t_m-\epsilon}_{t'=t}\K(t')\\
&&=\sum_{\rho_1,\rho_2,\dots \rho_m}
\bar w_{\rho_1,\rho_2\dots \rho_m}(t)
f_{\rho_1}(t_1) f_{\rho_2}(t_2)\dots f_{\rho_m}(t_m)\nn
\ea
with $t_{\hat n}=t+(\hat n-1)\epsilon$. The coefficients $\bar w$ are given (without index sums) by 
\be\label{N11}
\bar w_{\rho_1,\rho_2\dots \rho_m}(t)=\bar S_{\rho_m\rho_{m-1}}(t_m-\epsilon) \dots\bar S_{\rho_3\rho_2}\tee 
\bar S_{\rho_2\rho_1}(t).
\ee
(Note $\bar K(t+\epsilon,t)=\K(t),~\bar w_{\rho_1\rho_2}(t)=\bar S_{\rho_1\rho_2}(t)$.)

We generalize the boundary terms to an arbitrary expression of the sequences of initial and final occupation numbers $[n(\ti)]$ and $[n(t_f)]$,
\be\label{N12}
\bar b=\bar b(\ti,t_f)=f_\tau(\ti)b\tr f_\rho(t_f),
\ee
such that 
\be\label{N13}
w[n]=K[n]\bar b.
\ee
With $K[n]=\bar K(t_f, \ti)$ this expresses the weight function in terms of elements of the transfer matrix
\be\label{N14}
w[n]=\sum_{\rho_0\dots\rho_{G-1}}w_{\rho_0\dots \rho_{G-1}}
f_{\rho_0}(t_0)\dots f_{\rho_{G-1}}(t_{G-1})
\ee
with $t_{n'}=\ti +n'\epsilon$ and (without index sums)
\be\label{N15}
w_{\rho_0\dots \rho_{G-1}}=\bar S_{\rho_{G-1}\rho_{G-2}}\dots \bar S_{\rho_2\rho_1}
(\ti+\epsilon)\bar S_{\rho_1\rho_0}(\ti) b_{\rho_0\rho_{G-1}}.
\ee
Positivity of the weight function $w[n]$ requires that all coefficients obey $w_{\rho_0\dots \rho_{G-1}}\geq 0$. 

A positive weight distribution can be realized if all elements of $\bar S\tr (t)$ and all elements of $b\tr$ are positive semidefinite. In this case one has
\be\label{N26}
\K (t)=\exp \big\{-\cL(t)\big\}=\exp \big\{-f_\tau\tee M\tr (t)f_\rho (t)\big\},
\ee
with elements of the transfer matrix given by 
\be\label{N27}
\bar S\tr(t)=\exp \big\{-M\tr (t)\big\}.
\ee
This is the setting for generalized Ising models where $\cL(t)$ involves only Ising spins at two neighboring $t$-layers. We will mainly consider cases where $\bar S$ and $M$ do not depend on $t$. For constrained Ising models some elements $\bar S\tr$ vanish. This is realized by $M\tr \to\infty$ for the corresponding element of $M$. A positive matrix (e.g. all elements $\geq 0$) is not necessary for a positive weight - or probability distribution. Since eq. \eqref{N15} involves products of matrix elements it can be positive despite the appearance of negative elements in $\bar S$.

The functional integration is now performed easily. With eq. \eqref{7} one obtains the trace over matrix products
\be\label{N16}
Z={\rm tr}\big\{\bar S(t_f-\epsilon)\dots \bar S(\ti+\epsilon)\bar S(\ti)b\big\}.
\ee
In particular, for $\bar S$ independent on $t$ and periodic boundary conditions, $b=\bar S$, this yields the familiar expression of $Z$ in terms of the transfer matrix,
\be\label{N17}
Z={\rm tr} \{\bar S^G\}.
\ee

This extends to integrations over products of local evolution factors $\K(t')$. We define the evolution factor ${\cal S}(t_m,t)$ by integrating the product of $m-1$ subsequent $\K (t')$ factors over the intermediate configurations 
\be\label{N18}
{\cal S} (t_m,t)=
\left(\prod^{t_m-\epsilon}_{t'=t+\epsilon}\int dn(t')\right)
\bar K (t_m,t).
\ee
Thus ${\cal S}$ depends on two sets of local occupation numbers, namely $[n(t)]$ and $\big[n(t_m)\big]$. Employing eqs. \eqref{N10}, \eqref{N11} yields
\be\label{N19}
{\cal S}(t_m,t)=f_\tau(t_m)\bar U(t_m,t)_{\tau\rho}f_\rho (t),
\ee
with matrix product 
\be\label{N20}
\bar U(t_m,t)=\bar S(t_m-\epsilon)\dots \bar S\tee \bar S(t). 
\ee

\subsection{Observables and local bilinear operators}

A general local bilinear operator ${\cal B}(t)$ depends on two neighboring sets of local occupation numbers $[n\tee ]$ and $[n(t)]$,
\be\label{N21}
{\cal B}(t)=f_\tau\tee {B}_{\tau\rho}(t)f_\rho(t).
\ee
We define expectation values for products of bilinear operators as
\ba\label{26A}
&&\kl \B_B(t_2)\B_A(t_1)\kr =Z^{-1}\int \cD n{\cal S}(t_f,t_2+\epsilon)\B_B(t_2)\nn\\
&&\quad \times {\cal S}(t_2,t_1+\epsilon)\B_A(t_1){\cal S}(t_1,\ti)\bar b.
\ea
In other words, we replace in $K[n]$ the factors $\K(t_2)$ and $\K(t_1)$ by $\B_B(t_2)$ and $\B_A(t_1)$. In terms of matrices this yields
\ba\label{N22}
&&\kl {\cal B}_B(t_2){\cal B}_A(t_1)\kr =Z^{-1}{\rm tr}\big\{\bar U(t_f,t_2+\epsilon){B}_B(t_2)\nn\\
&&\qquad\times \bar U(t_2,t_1+\epsilon){B}_A(t_1)\bar U(t_1,\ti)b\big\}.
\ea
The generalization to time ordered products of several local bilinear operators is straightforward.

Local observables $(A(t)$ are represented by bilinears ${\cal B}(t)$ as
\be\label{N23}
A(t)=A_\tau(t)f_\tau(t)\to B\tr(t)=\bar S\tr (t)A_\rho(t),
\ee
such that eq. \eqref{N22} coincides with eq. \eqref{N6}. For example, the local occupation number $n_\alpha(t)$ corresponds to $A_\rho(t)=1$ if the sequence $\rho$ has a one at place $\alpha$, and $A_\rho(t)=0$ if there is a zero place $\alpha$. We can also express local observables $A\tee$ by the bilinear $B(t)$. In this case one has
\be\label{28A}
A\tee =A_\tau\tee f_\tau\tee \to B\tr(t)=A_\tau\tee \bar S\tr(t). 
\ee
At first sight the local observables are simpler objects and the reader may wonder about the usefulness of the local bilinear operators. We will see in sect. \ref{Grassmann observables} that these are the suitable objects for formulating the fermion-bit map on the level of observables. 

\subsection{Density matrix}

Even though not compulsory for the main argument of this paper and the formulation of the fermion-bit map it is instructive to express the expectation value of a single local bilinear in terms of the extended density matrix $\bar \rho(t)$
\be\label{N24}
\bar \rho\tr (t)= Z^{-1}\bl\bar U(t,\ti)b\bar U(t_f,t+\epsilon)\br_{\tau\rho}
\ee
as 
\be\label{N25}
\kl {\cal B}(t)\kr ={\rm tr}\big\{B(t)\bar \rho(t)\big\}.
\ee
This formula is familiar from quantum mechanics. 

The extended density matrix can be expressed by a functional integral over all $n(t')$ except $t'=t$ and $t'=t+\epsilon$,
\be\label{40A}
\hat\rho \T=f_\tau\T\bar\rho\tr \T f_\rho\tee,
\ee
with 
\be\label{40B}
\hat \rho\T=Z^{-1}\prod_{t'\neq t,t+\epsilon} dn(t')\hat w\T.
\ee
Here $\hat w\T$ is related to $w[n]$ by omitting in $K[n]$ the local factor $\K\T$,
\be\label{40C}
\hat w \T=\prod_{t'\neq t}\K(t')\bar b.
\ee
The normalization is given by 
\be\label{40D}
{\rm tr}\{\bar \rho\bar S\}=1,
\ee
and therefore differs from the one familiar in quantum mechanics if $\bar S\neq 1$. The ``missing factor'' $\bar S$ is included in the definition of the local bilinear operator. 

Arbitrary expectation values of products of occupation numbers at $t$ or $\tee$ can be computed by eq. \eqref{N25}. For example, the expectation value
\be\label{29A}
\kl \bl 1-n_\alpha\tee\br n_\beta(t)\kr =\sum\tr 
\bar A^{(\alpha)}_\tau\tee \bar S\tr (t)A^{(\beta)}_\rho(t)\bar \rho_{\rho\tau}(t)
\ee
involves the off-diagonal elements of $\bar \rho$ even for $\bar S\tr=\delta\tr$. Here the factors $\bar A^{(\alpha)}_\tau\tee$ and $A^{(\beta)}_\rho$ equal one or zero and effectively restrict the sums. Denoting $\rho$ by a sequence $[\rho_\gamma]$ of numbers one or zero, and similar for $\tau$, one has
\be\label{29B}
\kl \bl 1-n_\alpha\tee\br n_\beta(t)\kr=\sum_{\tau,\tau_\alpha=0}\sum_{\rho,\rho_\beta=1}
\bar S\tr(t)\bar \rho_{\rho\tau}(t).
\ee
In particular, for $\beta=\alpha$ only off-diagonal elements of $B\tr$ contribute, (no index sums)
\be\label{42A}
B^{(\alpha)}\tr=\bar A^{(\alpha)}_\tau\tee \bar S\tr\T A^\alpha_\rho \T. 
\ee
The observable $(1-n_\alpha\tee \br n_\alpha\T$ is represented by an off-diagonal local bilinear operator. Correspondingly, its expectation value involves the off-diagonal elements of the extended density matrix.

For local observables \eqref{N23} one may use the restricted local density matrix $\rho'(t)$ which is related to $\bar \rho(t)$ by 
\be\label{30A}
\rho'(t)=\bar \rho(t)\bar U(t+\epsilon,t)=\bar U(t,\ti)b\bar U(t_f,t)=\bar \rho(t)\bar S(t). 
\ee
The expectation value reads
\be\label{30B}
\kl A(t)\kr ={\rm tr}\big\{A'(t)\rho'(t)\big\},
\ee
where 
\be\label{30C}
A'\tr (t)=A_\tau(t)\delta\tr.
\ee
In this case only diagonal elements of $\rho'$ contribute to eq. \eqref{30B}. The normalization is ${\rm tr} \rho'=1$. For invertible $\bar S$ this can be extended to arbitrary bilinear operators by associating to $B$ an operator
\be\label{YY1}
A'=\bar S^{-1}B,
\ee
such that eq. \eqref{N25} becomes 
\be\label{YY2}
\kl \B\T\kr ={\rm tr}\big\{A'\T\rho'\T\big\}.
\ee
The restricted density matrix $\rho'$ contains sufficient information for the computation of expectation values of arbitrary observables constructed from occupation numbers on two neighboring $t$-layers.

\section{Grassmann functional integral}
\label{Grassmann functional integral}

The probabilistic models for occupation numbers can be interpreted in a fermionic language. If $n_\gamma(t)=1$ a fermion of species $\gamma$ is present at $t$, while for $n_\gamma(t)=0$ it is absent. If for a given $t$ more than one occupation number $n_\gamma(t)$ differs from zero, more than one fermion is present at $t$. Our models therefore describe multi-fermion systems as well. This extends to a higher dimensional setting - for $n_\alpha(\vec x,t)=1$ a fermion of species $\alpha$ is present at position $(\vec x,t)$. We will see that for suitable models one can associate $t$ with time, such that at fermion $\alpha$ is present at time $t$ at position $\vec x$. 

In quantum field theory fermions are usually described by functional integrals over Grassmann variables. In the present and next sections we discuss a general map between the ``occupation number basis'' of the preceding section and a Grassmann formulation. This map establishes the equivalence of fermionic models with suitable models of Ising spins. Only specific fermionic models will correspond to a positive probability distribution $p[n]$, while for the general case one obtains an indefinite weight distribution $w[n]$. In the fermionic representation we use Grassmann variables $\psi_\gamma(t)$ instead of occupation numbers $n_\gamma(t)$ as the basic variables. The number of variables is the same in both formulations. The $2^M$ basis functions $f_\tau(t)$ can be mapped to $2^M$ basis functions $g_\tau(t)$ of the Grassmann algebra constructed from $\psi_\gamma(t)$. In the fermionic language we will deal with weight distributions $\tilde w[\psi]$ that are functionals of Grassmann variables at all $t$. 

The basis of the map between the Grassmann basis and the occupation number basis is simple. Consider a model for Ising spins characterized by $w[n]$ and a functional integral for Grassmann variables $\psi_\gamma(t)$. They are equivalent if both lead to the same transfer matrices $\bar S(t)$, and if a suitable fermionic representation of observables can be found such that eq. \eqref{N22} and generalizations for multiple products hold. Then all expectation values of products of observables are the same in both models. They describe the same physics in different languages. If the transfer matrices $\bar S(t)$ define the model uniquely, the map between the occupation number basis and the Grassmann or fermion basis is invertible. In the present section we establish the map between the weight functions $w[n]$ and $\tilde w[\psi]$. This will be completed in the next section by the map between local bilinear operators.

\subsection{Map of basis functions and wave functions}

We want to establish a map between the basis functions $f_\tau(t)$, which depend on the local occupation numbers $n_\gamma(t)$, and the basis functions of the Grassmann algebra $g_\tau(t)$, which are polynomials of the local Grassmann variables $\psi_\gamma(t)$. The basis functions $f_\tau$ involve $M$ powers of factors $a_\gamma$ which are either given by $n_\gamma$ or $(1-n_\gamma)$,
\be\label{B1a}
f_\tau=\prod_\gamma a_\gamma.
\ee
We can keep this structure in the fermionic formulation constructing basis elements of the Grassmann algebra as,
\be\label{B2a}
g_\tau=\tilde s_\tau\prod_\gamma\tilde a_\gamma,
\ee
where $\tilde a_\gamma=\psi_\gamma$ if $a_\gamma=(1-n_\gamma)$ and $\tilde a_\gamma=1$ if $a_\gamma=n_\gamma$. The basis elements $g_\tau$ are products of up to $M$ Grassmann variables. They span the complete local Grassmann algebra. Since Grassmann variables anticommute eq. \eqref{B2a} is only well defined if one specifies an order of the product. We will use the order of the labels $\gamma=1\dots M$, and take always products of $\psi_\gamma$ with smaller $\gamma$ to the left. It is often convenient to introduce a sign factor $\tilde s_\tau=\pm 1$ which is left arbitrary for the moment. For a given choice of signs $\tilde s_\tau$ the map between $f_\tau$ and $g_\tau$ is invertible. 

The anticommuting properties of Grassmann variables will force us to keep track of signs carefully. This will make the establishment of the fermion-bit map somewhat cumbersome. Once the map is established, however, all issues of signs are taken into account automatically by the properties of the Grassmann variables. 

An arbitrary element of the local Grassmann algebra is called a Grassmann wave function $g(t)$, 
\be\label{B3a}
g(t)=q_\tau(t)g_\tau(t).
\ee
Similarly, we may employ a wave function $f(t)$ in the occupation number basis,
\be\label{B4a}
f(t)=q_\tau(t)f_\tau(t),
\ee
using the same coefficients $q_\tau(t)$. The map between basis functions $f_\tau$ and $g_\tau$ extends to the map between wave functions $f(t)$ and $g(t)$. These wave functions and the corresponding map are discussed in ref. \cite{IMCWF}. The present paper will not use wave functions. We rather establish the map on the more fundamental level of weight distributions. This will reveal an important modulo two property related to particle-hole conjugation. 

\subsection{Particle-hole conjugation}

The map between $f_\tau$ and $g_\tau$ is not unique. For example, we may define conjugate basis functions $g^c_\tau$ as
\be\label{B5a}
g^c_\tau=\tilde s^c_\tau\prod_\gamma\tilde b_\gamma,
\ee
with $\tilde b_\gamma=1$ if $a_\gamma=(1-n_\gamma)$, and $\tilde b_\gamma=\psi_\gamma$ if $a_\gamma=n_\gamma$. The induced linear map between $g_\tau$ and $g^c_\tau$ exchanges the association of basis elements to occupied and unoccupied states. It may therefore be called a particle-hole conjugation, represented by a conjugation matrix ${\cal C}$,
\be\label{B6a}
g^c_\tau={\cal C}_{\tau\rho}g_\rho.
\ee

Different choices of the signs $\tilde s^c_\tau$ define different versions of conjugate basis functions. We will employ a particular particle-hole conjugation \eqref{B5a}, \eqref{B6a} where the signs $\tilde s^c_\tau$ and $\tilde s_\tau$ are related by the defining relation
\be\label{B7a}
g^c_\tau[\varphi]=\int d\psi \exp (\varphi_\gamma\psi_\gamma) g_\tau[\psi].
\ee
Here the order of the Grassmann integrations is specified as
\be\label{B8a}
\int d\psi =\int d\psi_M\dots \it d\psi_1,
\ee
e.g. the larger $\gamma$ to the left. We consider even $M$ such that $\int d\psi$ commutes with Grassmann variables. With 
\be\label{B9a}
\int d\psi \exp (\varphi_\gamma\psi_\gamma)=\prod_\gamma\int d\psi_\gamma(1+\varphi_\gamma\psi_\gamma)
\ee
it is easy to see that every factor $\psi_\gamma$ in $g$ will be mapped to a factor $1$ in $g^c$, while every factor $1$ in $g$ results in a factor $\varphi_\gamma$ in $g^c$. This is precisely the action of ${\cal C}$, whereby the definition \eqref{B7a} fixes the relative signs between elements $g_\tau$ and $g^c_\tau$. 

Repeating the particle-hole conjugation yields
\ba\label{B10a}
g^{cc}_\tau[\chi]&=&\int d\varphi\exp (\chi_\gamma\varphi_\gamma)g^c_\tau(\varphi)=\nn\\
&=&\int d\varphi d\psi\exp \big[\varphi_\gamma(\psi_\gamma-\chi_\gamma)\big]g_\tau(\psi)\nn\\
&=&\eta_M  g_\tau(\chi)=({\cal C}^2)\tr g_\rho(\chi),
\ea
with $\eta_M =-1$ for $M=1,2$ mod $4$ and $\eta_M =1$ for $M=3,4$ mod $4$. One  establishes
\be\label{B11a}
{\cal C}^2=\eta_M ,~{\cal C}^4=1. 
\ee

We next collect several identities for products of Grassmann basis elements that will be employed for the map between weight functions. As a useful identity one may use 
\be\label{B12a}
\exp (\varphi_\gamma\psi_\gamma)=\prod_\gamma(1+\varphi_\gamma\psi_\gamma)=\sum_\tau \epsilon_\tau g_\tau (\varphi) g_\tau(\psi),
\ee
with $\epsilon_\tau=\pm 1$. If we denote by $m_\tau$ the number of $\psi$-factors in $g_\tau(\psi)$, one finds $\epsilon_\tau=1$ for $m_\tau=0,1$ mod $4$ and $\epsilon_\tau=-1$ for $m_\tau=2,3$ mod $4$. This can be cast into the form
\be\label{44A}
\epsilon_\tau=(-1)^{\frac{m_\tau(m_\tau-1)}{2}}.
\ee
We also can write, with $\epsilon^c_\tau=\pm 1$, 
\be\label{B13a}
\exp (\varphi_\gamma\psi_\gamma)=\sum_\tau \epsilon^c_\tau g^c_\tau (\varphi) g^c_\tau(\psi).
\ee
Now $\epsilon^c_\tau=1$ for $m^c_\tau=M-m_\tau=0,1$ mod $4$ and $\epsilon^c_\tau=-1$ for $M-m_\tau=2,3$ mod $4$. For $M=0$ mod $4$ one has $\epsilon^c_\tau=(-1)^{m_\tau}\epsilon_\tau$, while for $M=2$ mod $4$ one finds $\epsilon^c_\tau=(-1)^{m_\tau+1}\epsilon_\tau$, thus
\be\label{B14a}
\epsilon^c_\tau=(-1)^{m_\tau}\eta (M)\epsilon_\tau. 
\ee

We next observe the relations
\ba\label{B15a}
\int d\psi g_\tau(\psi)g^c_\rho(\psi)&=&\sigma_\tau\delta\tr,\nn\\
\int d\psi g^c_\tau (\psi)g_\rho(\psi)&=&(-1)^{m_\tau}\sigma_\tau\delta\tr,
\ea
with $\sigma_\tau=\pm 1$. The signs $\sigma_\tau$ can be related to $\epsilon_\tau$ by use of (no sum over $\tau$)
\ba\label{B16a}
g_\tau(\psi)&=&{\cal C}^{-1}\tr g^c_\rho(\psi)=\eta_M {\cal C}\tr g^c_\rho (\psi)\nn\\
&=&\eta_M \int d\varphi\exp (\psi_\gamma\varphi_\gamma)g^c_\tau(\varphi)\nn\\
&=&\eta_M \int d\varphi \sum_\rho\epsilon_\rho g_\rho(\psi) g_\rho(\varphi) g^c_\tau(\varphi)\nn\\
&=&\eta_M \epsilon_\tau\sigma_\tau g_\tau(\psi),
\ea
or 
\be\label{B17a}
\sigma_\tau=\eta_M \epsilon_\tau.
\ee

We further define (no sum over $\tau$)
\be\label{B18a}
g'_\tau(\psi)=\epsilon_\tau g_\tau(\psi).
\ee
This allows us to write
\be\label{B19a}
\exp (\varphi_\gamma\psi_\gamma)=g_\tau(\varphi)g'_\tau(\psi),
\ee
and 
\be\label{B20a}
\int d\psi g'_\tau(\psi)g^c_\rho(\psi)=\eta_M \delta\tr.
\ee

Conjugate Grassmann elements $\bar g_\tau$ are defined by 
\be\label{B23a}
\bar g_\tau =\epsilon^c_\tau g^c_\tau =(-1)^{m_\tau}\epsilon_\tau \eta (M)g^c_\tau.
\ee
With this definition eq. \eqref{B13a} reads
\be\label{XYZ}
\exp (\varphi_\gamma\psi_\gamma)=g^{c}_\tau(\varphi)\bar g_\tau(\psi).
\ee
Conjugate Grassmann elements obey the integration rule
\ba\label{B25a}
\int d\psi \bar g_\rho (\psi)g_\tau(\psi)&=&\delta_{\rho\tau}.
\ea
The relations \eqref{B19a}, \eqref{B20a}, \eqref{XYZ}, \eqref{B25a} will be used repeatedly for the map between the weight functions $w[n]$ and $\tilde w[\psi]$. In particular, eqs. \eqref{B20a} and \eqref{B25a} play the role of eq. \eqref{N7}. Since they involve both basis elements and conjugate elements our formulation will employ both relations in an alternating fashion. 

\subsection{Evolution factor and transfer matrix}
We next establish the map between the evolution operators which define the transfer matrix. We use two alternating maps for alternating ``even and odd positions''. This is necessary because of the modulo two property related to particle-hole conjugation. The evolution factor $\K(t)$ in the occupation number basis can be mapped to a Grassmann evolution factor
\be\label{B21a}
\tilde \K(t)=g_\tau\tee \bar S\tr(t)g'_\rho(t),
\ee
with $g_\rho(t)=g_\rho\bl \psi(t)\br$. Another map defines 
\be\label{B22a}
\tilde \K^c(t)=g^{c}_\tau \tee \bar S\tr (t)\bar g_\rho(t).
\ee
Here $\bar S\tr\T$ is the same transfer matrix as in the occupation number basis. We will employ $\K\T$ for odd $t$ and $\tilde \K^c\T$ for even $t$. Even and odd $t$ correspond to even and odd $n'$ in $t_{n'}=\ti +n'\epsilon$. 

The product of two subsequent evolution factors obeys 
\be\label{M1a}
\tilde \K^c\tee \tilde \K(t)=g^{c}_\tau(t+2\epsilon)\bar S_{\tau\alpha}\tee F_{\alpha\beta}\tee \bar S_{\beta\rho}(t)g'_\rho(t),
\ee
with 
\be\label{M2a}
F_{\alpha\beta}\T=\bar g_\alpha\T g_\beta\T,~\int d\psi\T F_{\alpha\beta}\T=\delta_{\alpha\beta}.
\ee
Similarly, one has
\be\label{M3a}
\tilde\K\T\tilde\K^c(t-\epsilon) =g_\tau\tee \bar S_{\tau\alpha}\T\tilde F_{\alpha\beta}\T\bar S_{\beta\rho}(t-\epsilon) \bar g_\rho(t-\epsilon),
\ee
with 
\be\label{M4a}
\tilde F_{\alpha\beta}\T=g'_\alpha\T g^{c}_\beta\T,~\int d\psi \T\tilde F_{\alpha\beta}\T=
\eta_M \delta_{\alpha\beta}.
\ee
Integration over intermediate Grassmann variables results in matrix multiplication of transfer matrices
\ba\label{69AA}
&&\int d\psi \T \tilde\K^c\T \tilde\K(t-\epsilon)=g^{c}_\tau(t+\epsilon)\nn\\
&&\qquad\times\big[\bar S\T \bar S(t-\epsilon)\big]\tr g'_\rho(t-\epsilon),
\ea
and
\ba\label{M5a}
&&\int d\psi\tee \tilde \K\tee \tilde \K^c(t) =\eta_M g_\tau\teee\nn\\
&&\qquad\times \big[\bar S\tee\bar S(t)\big]\tr \bar g_\rho(t).
\ea
This extends to longer sequences of alternating $\tilde \K$ and $\tilde \K^c$ factors, as
\ba\label{M6a}
&&\tilde \K(t+\epsilon)\tilde \K^c\T \tilde \K(t-\epsilon)=g_\tau(t+2\epsilon)\bar S_{\tau\alpha}(t+\epsilon)\nn\\
&&\quad \times \tilde F_{\alpha\beta}(t+\epsilon)\bar S_{\beta\gamma}\T F_{\gamma\delta}(t)\bar S_{\delta\rho}(t-\epsilon) g'_\rho(t-\epsilon),
\ea
with 
\ba\label{M7a}
&&\int d\psi (t+\epsilon)d\psi\T \tilde \K (t+\epsilon)\tilde \K^c\T \tilde \K-\epsilon(t-\epsilon)\\
&&=\eta_M g_\tau(t+2\epsilon)\big[\bar S(t+\epsilon)\bar S\T \bar S(t-\epsilon)\big]\tr g'_\rho(t-\epsilon).\nn
\ea
The definitions \eqref{B21a}, \eqref{B22a} therefore realize the crucial matrix product of transfer matrices that appears in expressions as eqs. \eqref{N20}, \eqref{N22}.

\subsection{Grassmann weight function}

The map from the occupation number basis to the Grassmann basis associates to each chain of $\K(t')$ factors an alternating chain of $\tilde \K^c$ and $\tilde \K$ factors. By convention we start at $\ti$ with $\tilde\K^c(\ti)$ and associate to $\K(\ti+n'\epsilon)$ a Grassmann evolution factor $\tilde\K^c(\ti+n'\epsilon)$ if $n'$ is even, and $\tilde \K(\ti+n'\epsilon)$ if $n'$ is odd. The transfer matrices $\bar S\tr \T$ used in the representations \eqref{N8}, \eqref{B21a}, \eqref{B22a} are the same. The factor $K[n]$ in eq. \eqref{N2} is mapped to the corresponding chain $\tilde K[\psi]$ of alternating $\tilde \K$ and $\tilde \K^c$ factors.

In order to define the map of the weight distribution $w[n]$ to a weight function $\tilde w[\psi]$ in the Grassmann basis we have to define the map for the boundary term $\bar b$. Let us consider an even number $G$ of $t$-points such that the number of $\K$-factors in $K[\psi]$ is odd. We define
\be\label{M8a}
\tilde b[\psi]=\tilde b\big[\psi(\ti),~\psi(t_f)\big]=\eta^{\frac G2}_Mg_\tau(\ti)b\tr g'_\rho(t_f).
\ee
This yields the map $w[n]\to \tilde w[\psi]$, with 
\be\label{M9a}
\tilde w[\psi]=K[\psi]\tilde b[\psi].
\ee
By virtue of the definitions \eqref{B21a}, \eqref{B22a} and the product rules \eqref{M1a}-\eqref{M4a} the Grassmann weight function $\tilde w[\psi]$ is a closed chain of factors
\ba\label{M10a}
\tilde w[\psi]&=&\tilde F_{\rho_{G-1}\tau_{G-1}}(t_f)\bar S_{\tau_{G-1}\tau_{G-2}}(t_f-\epsilon)\dots F_{\tau_2\rho_2}(t+2\epsilon)\nn\\
&&\times \bar S_{\rho_2\rho_1}\tee \tilde  F_{\rho_1\tau_1}\tee\bar S_{\tau_1\tau_0}\T F_{\tau_0\rho_0}\T b_{\rho_0\rho_{G-1}}.\nn\\
\ea
Comparison with eq. \eqref{N14} shows that the product of basis functions $f_\tau$ is now replaced by a product of Grassmann bilinears $F,\tilde F$. Correspondingly, the coefficients are again products of elements of transfer matrices, now with the double number of indices. The role of the integration rule \eqref{6} is played by the identities \eqref{M2a}, \eqref{M4a}. 

For the Grassmann functional integral one obtains
\be\label{A4-5}
\int \cD \psi\tilde w[\psi]={\rm tr}
\big\{\bar S(t_f-\epsilon)\dots \bar S(\ti) b\big\}=Z.
\ee
This is the same formula as eq. \eqref{N16} in the occupation number basis. The two formulations are equivalent in the sense that the partition function is the same. Since both $w[n]$ and $\tilde w[\psi]$ are uniquely specified by the transfer matrices $\bar S(t')$ and the boundary factor $b$, the map between the two formulations is an isomorphism. Every quasilocal Grassmann functional integral is mapped in this way to a weight distribution of Ising spins, and vice versa. What remains is a specification of the observables in the Grassmann basis. 

\section{Grassmann observables}
\label{Grassmann observables}

In this section we will map the local bilinear operators $\B\T$ in the occupation number basis to corresponding bilinear Grassmann operators $\tilde \B\T$ and $\tilde \B^c\T$ in the fermionic basis. According to the modulo two property of the particle-hole conjugation we employ $\tilde \B\T$ for odd $t$ and $\tilde \B^c\T$ for even $t$. The definitions will be chosen such that the Grassmann functional integral yields expectation values given by generalizations of eq. \eqref{N22}. The map between observables will complete the general fermion-bit map.

\subsection{Bilinear Grassmann operators}

Bilinear Grassmann operators $\tilde \B\T$ and $\tilde \B^c\T$ involve Grassmann variables at two neighboring positions $t$ and $t+\epsilon$. According to the alternating sequences of $\tilde \K\T$ and $\tilde \K^c\T$ we define for odd $t$,
\be\label{G1}
\tilde \B\T=g_\tau\tee B\tr\T g'_\rho\T,
\ee
while we use for even $t$,
\be\label{G2}
\tilde \B^c\T=g^{c}_\tau\tee B\tr \T \bar g_\rho\T.
\ee
Here we use the same matrices $B\tr\T$ as for the bilinear operators \eqref{N21} in the occupation number basis. 

The expectation value of a product of bilinear operators employs in $\tilde w[\psi]$ the Grassmann bilinears $\tilde \B\T$ instead of $\tilde \K\T$, and $\tilde \B^c\T$ instead of $\tilde \K^c\T$, and subsequently multiplies by $Z^{-1}$. For example, for $t_1$ and $t_2$ odd the expectation value \eqref{26A} becomes in the Grassmann basis
\ba\label{G3}
&&\kl \B_B(t_2)\B_A(t_1)\kr =Z^{-1}\int \cD\psi \tilde {\cal S}(t_f,t_2+\epsilon)\tilde \B_B(t_2)\nn\\
&&\qquad\times \tilde {\cal S}(t_2,t_1-\epsilon)\tilde \B_A(t_1)\tilde {\cal S}(t_1,\ti)\tilde b.
\ea
Here $\tilde{\cal S}(t_m,t)$ is constructed in analogy to eq. \eqref{N18} by a product of alternating factors of $\tilde \K^c$ and $\tilde \K$, with intermediate Grassmann variables integrated out. Insertion of the definitions \eqref{G1}, \eqref{G2} yields the matrix product \eqref{N22}. We conclude that the definition of bilinear Grassmann operators \eqref{G1}, \eqref{G2}, together with the rule \eqref{G3}, yields the same expectation values for all products of local bilinear observables. This completes the map between the occupation number basis and the Grassmann basis. 

\subsection{Local occupation numbers in the Grassmann}

~{\bf basis}

What remains to be done is the computation of the explicit form of the Grassmann bilinears $\tilde \B$ and $\tilde \B^c$ for products of occupation numbers. This is based on the general definitions \eqref{G1}, \eqref{G2}. For the local occupation number $n_\alpha\T$ the associated bilinear operator corresponds to matrices $B^{(\alpha)}\T$,
\be\label{GB1}
B^{(\alpha)}\tr\T=\left\{
\begin{array}{ccc}
\bar S\tr\T&{\rm for}&\rho_\alpha=1\\
0&{\rm for}&\rho_\alpha=0
\end{array}.
\right.
\ee
Here we consider $\rho$ as a sequence of $M$ numbers $\rho_\gamma$ taking values one or zero, $\rho=(\rho_1\dots \rho_\alpha\dots \rho_M$). Thus $\tilde \B^{(\alpha)}$ is constructed similar to $\tilde \K$, but with a sum over $\rho$ restricted to those terms where $\rho_\alpha=1$,
\be\label{HB1}
\tilde \B^{(\alpha)}\T=\sum_\tau\sum_{\rho,\rho_\alpha=1}
g_\tau\tee \bar S\tr \T g'_\rho\T. 
\ee
In other words, $\tilde \B^{(\alpha)}\T$ obtains from $\tilde \K\T$ by eliminating all terms that have a factor $\psi_\alpha\T$. For $\tilde \B^{c(\alpha)}\T$ the role of particles and holes is exchanged. Now $\tilde \B^{c(\alpha)}\T$ obtains from $\tilde \K^c\T$ by eliminating all terms that do not contain a factor $\psi_\alpha\T$,
\be\label{HB2}
\tilde \B^{c(\alpha)}\T=\sum_\tau\sum_{\rho,\rho_\alpha=1}g^{c}\tee \bar S\tr \T\bar g_\rho\T.
\ee

Eqs. \eqref{HB1} and \eqref{HB2} give a conceptually very simple explicit definition of the bilinear Grassmann operators corresponding to the local occupation numbers $n_\alpha\T$. One modifies the evolution factors such that only states with $n_\alpha\T=1$ ``propagate''. Similarly, for the observable $1-n_\alpha\T$ only states with $n_\alpha\T=0$ propagate. This is easily extended to products of local particle numbers. For the product $n_\alpha\T n_\beta\T$ one further restricts the sum in eq. \eqref{HB1},
\be\label{HB3}
\tilde \B^{(\alpha,\beta}\T=\sum_\tau\sum_{\rho,\rho_\alpha=\rho_\beta=1}
g_\tau \tee \bar S\tr\T g'_\rho\T,
\ee
and similar for higher products. 

We observe that the bilinear operators $\B\T$ can also be used to compute expectation values of occupation numbers at $t+\epsilon$. The observable $n_\alpha\tee$ is now represented by restricting the sum over $\tau$ to $\tau_\alpha=1$,
\be\label{HB4}
n_\alpha\tee \to \B^{(\bar \alpha)}\T=\sum_{\tau,\tau_\alpha=1}\sum_\rho g_\tau\tee \bar S\tr g'_\rho\T. 
\ee
Similarly, the product $n_\alpha\tee n_\beta\T$ is mapped to 
\be\label{HB5}
n_\alpha\tee n_\beta\T\to \tilde \B^{(\bar\alpha,\beta)}\T=\sum_{\tau,\tau_\alpha=1}\sum_{\rho,\rho_\beta=1}g_\tau\tee \bar S\tr g'_\rho\T. 
\ee
The bilinear operators $\tilde \B\T$ can therefore represent arbitrary products of occupation numbers at $t$ or $t+\epsilon$. This holds analogously for $\tilde \B^c\T$.

Products of occupation numbers at arbitrary $t$ can be computed by replacements $\tilde \K\to \tilde \B,~\tilde \K^c\to \tilde \B^c$ at several positions $t$, according to eq. \eqref{G3}. This provides for a complete explicit representation of all observables constructed from occupation numbers in the Grassmann basis. We note that an observable as $n_\alpha\tee$ at odd $t$ can be represented equivalently by $\tilde \B\T$ or $\tilde \B^c\tee$, and for even $t$ by $\tilde \B^c\T$ or $\tilde \B\tee$. 

\subsection{Annihilation and creation operators}

One would like to have a simple formal expression for the elimination of terms in the sums for $\tilde \B$ or $\tilde \B^c$. This can be achieved by the use of Grassmann derivatives. For this purpose we note the identity (no sum over $\alpha$)
\be\label{GB2}
\frac{\partial}{\partial \psi_\alpha}\psi_\alpha g_\rho=
\left\{
\begin{array}{ccc}
g_\rho&{\rm for}&\rho_\alpha=1\\
0&{\rm for}&\rho_\alpha=0
\end{array}.
\right.
\ee
Here we recall that the sequence of Grassmann variables $g_\rho$ contains a factor $\psi_\alpha$ if $\rho_\alpha=0$, and no such factor if $\rho_\alpha=1$. For odd $t$ the matrix \eqref{GB1} is therefore realized by 
\be\label{GB3}
\tilde {\cal N}_\alpha\T=\tilde \B^{(\alpha)}=
\frac{\partial}{\partial \psi_\alpha\T}\psi_\alpha\T\tilde \K\T.
\ee
For even $t$ we use the relation
\be\label{GB4}
\psi_\alpha\frac{\partial}{\partial \psi_\alpha}\bar g'_\rho=\left\{
\begin{array}{ccc}
\bar g'_\rho&{\rm for}&\rho_\alpha=1\\
0&{\rm for}&\rho_\alpha=0
\end{array}.
\right.
\ee
This yields
\be\label{GB5}
\tilde \cN^c_\alpha\T=\psi_\alpha\T\frac{\partial}{\partial \psi_\alpha\T}\tilde \K^c\T.
\ee
The alternating factors $\tilde \K\T$ and $\tilde \K^c\T$ in the construction of $\tilde w[\psi]$ are reflected in the alternating choices of the bilinears $\tilde \cN_\alpha\T$ and $\tilde \cN^c_\alpha\T$. 

The factor $\cN_\alpha\T$ replaces a factor $\K\T$ in the chain for $\tilde w[\psi]$. Since it involves a factor $\tilde\K \T$ we can also express expectation values by an insertion of the factors
\be\label{JB2}
\tilde n_\alpha\T=\frac{\partial}{\partial\psi_\alpha\T}\psi_\alpha\T,~\tilde n^c_\alpha=\psi_\alpha\T
\frac{\partial}{\partial \psi_\alpha\T},
\ee
at appropriate places. This extends to products of local observables and also covers occupation numbers at $t+\epsilon$. For example, the Grassmann operator associated to $n_\alpha\tee n_\beta\T n_\gamma\T$ is expressed as (no sum over indices)
\be\label{JB1}
n_\alpha\tee n_\beta\T n_\gamma\T\to \B^{(\bar \alpha,\beta,\gamma)}=\tilde n_\alpha\tee \tilde n_\beta\T\tilde n_\gamma\tilde\T \K\T.
\ee
For products of observables at different $t$ we can use appropriate products of bilinear Grassmann operators at the corresponding positions. 

We mainly use $\tilde \K\T$ and $\tilde \K^c\T$ factors that only contain terms with an even number of Grassmann variables. Then they commute among themselves and with all $\tilde \B$ and $\tilde \B^c$ factors. Still, the place of the factors $\tilde \B,\tilde \B^c$ in the string \eqref{G3} is fixed since they replace factors $\tilde \K,\tilde \K^c$ at the corresponding positions. This fixes the place of the insertion of the derivative operators $\tilde n_\alpha\T$ or $\tilde n^c_\alpha\T$. 

In eqs. \eqref{GB3} or \eqref{GB5} the operators $\tilde n_\alpha\T$ or $\tilde n^c_\alpha\T$ act on $\tilde \K\T$ or $\tilde \K^c\T$ within the corresponding expression for $\tilde \B\T$ or $\tilde \B^c\T$. One may also wish to have a formulation where Grassmann operators are inserted in the functional integral, with derivatives $\partial/\partial\psi_\alpha\T$ acting on all variables to the right in the chain until $\tilde b(\ti,t_f)$. Assume $t$ to be odd. For occupation numbers $n_\alpha\tee$ we insert in $\tilde w[\psi]$ a factor $\tilde n_\alpha\tee$ between $\tilde \K\tee$ and $\tilde \K\T$,while for $n_\alpha\T$ one inserts $\tilde n^c_\alpha\T$  between $\tilde \K\T$ and $\tilde \K\tee$. This may be seen  by observing that $\tilde n_\alpha\tee$ in $\tilde \K\T$ stands already to the left of all other variables $\psi\tee$, and the same holds for $\tilde n^c_\alpha\T$ in $\tilde \K^c(t-\epsilon)$. For $\tilde n_\alpha\T$ in $\tilde \K\T$ one may use
\ba\label{JB3}
\int d\psi\T\tilde n_\alpha\T\tilde \K\T\tilde \K^c(t-\epsilon)=
\int d\psi\T\tilde\K\T\tilde n^c_\alpha\T\tilde \K(t-\epsilon),\nn\\
\ea
which leads to the same result. Expectation values of occupation numbers can therefore be computed by insertion of $\tilde n_\alpha\T$ or $\tilde n^c_\alpha\T$, depending on the place of the insertion. 

We may define annihilation operators $a_\gamma\T$ and creation operators $a^\dagger_\gamma\T$ by 
\ba\label{JB4}
a_\gamma\T&=&
\left\{
\begin{array}{cccc}
\psi_\gamma&{\rm for}&t&{\rm even}\\
\frac{\partial}{\partial\psi_\gamma\T}&{\rm for}&t&{\rm odd}
\end{array}.
\right.\nn\\
a^\dagger_\gamma\T&=&
\left\{
\begin{array}{cccc}
\frac{\partial}{\partial\psi_\gamma\T}&{\rm for}&t&{\rm even}\\
\psi_\gamma\T&{\rm for}&t&{\rm odd}
\end{array}.
\right.
\ea
with (no index sums)
\be\label{JB5}
a^\dagger_\gamma\T a_\gamma\T=
\left\{
\begin{array}{cccc}
\tilde n_\gamma\T&{\rm for}&t&{\rm even}\\
\tilde n^c_\gamma\T&{\rm for}&t&{\rm odd}
\end{array}.
\right.
\ee
Observables involving factors of $n_\alpha\T$ can then be realized by inserting $a^\dagger_\alpha\T a_\alpha\T$ at the appropriate places. This is a standard setting for occupation numbers for the quantum field theory description of fermions. We conclude that the map of local bilinear operators in the occupation number basis to the Grassmann basis by eqs. \eqref{G1}, \eqref{G2} results in a rather simple and well known rule for the computation of expectation values for occupation numbers in quantum field theories for fermions, formulated as Grassmann functional integrals. 

\section{Simple fermion models}
\label{Simple fermion models}

The map between evolution factors in the occupation number basis $\K\T$ and fermion basis $\tilde \K\T,\tilde \K^c \T$ establishes a simple general structure for the fermion-bit map. In particular, it proves that such a map is always possible. It also ensures locality in $t$. For a small number of Grassmann variables it can be used very directly for finding the occupation number representation for a given fermion model, or inversely the fermion representation for a given generalized Ising model. For large $M$ the direct ``brute force construction'' becomes rapidly cumbersome since the number of basis elements is $2^M$, and the transfer matrix therefore a $2^M\times 2^M$ matrix. For a practical use of the fermion-bit map for large $M$, as for the example of models in two or more dimensions, it is important to understand better the basic structures of the fermion-bit map. In this section we discuss simple fermionic models that can be solved exactly. They translate to equivalent models for Ising spins. 

\subsection{Trivial model}

Consider the Grassmann evolution factors
\be\label{SF1}
\tilde \K\T=\exp \big\{\psi_\gamma\tee \psi_\gamma\T\big\}.
\ee
By virtue of eqs. \eqref{B19a}, \eqref{B21a} one infers for odd $t$
\be\label{SF2}
\bar S\tr\T=\delta\tr.
\ee
Let us employ the same Grassmann evolution factor for even and odd $t$
\be\label{SF3}
\tilde \K^c\T=\tilde \K\T. 
\ee
The relation \eqref{SF2} extends to even $t$ by virtue of eqs. \eqref{XYZ}, \eqref{B22a}. The transfer matrix for all $t$ is the unit matrix. The weight function \eqref{N14}, \eqref{N15} is a chain of unit matrices and $b$, with $Z=$tr $b$. By a suitable normalization of $b$ one has $Z=1$. The representation in terms of Ising spins is given by eq. \eqref{N26} with 
\be\label{SF4}
M\tr=-\ln \delta\tr.
\ee
This is a type of constrained Ising model \cite{CWIT} where for $\tau\neq \rho$ the matrix element $M\tr$ diverges. (One may realize this as a limiting procedure of corresponding $M\tr$ growing very large.) The diagonal elements of $M$ are zero.

Despite its trivial character it is worthwhile to discuss a few features of this model, since it may be the expansion point for neighboring models where the transfer matrix shows only small deviations from the unit matrix. 

We first note that the trivial model is a perfect static memory material \cite{CWIT}. Information at the ``initial boundary'' at $\ti$ is transported to the ``final boundary'' at $t_f$ without any loss. Consider a product form \eqref{N12} of the boundary conditions, where ``initial conditions'' are set by $f_{in}$, while no bias distinguishes possible states at $t_f$, e.g. $\bar f_f=1$. We expand
\be\label{SF5}
f_{in}=\tilde q_\rho (\ti)f_\rho(\ti),~b\tr =\tilde q_\rho(\ti),
\ee
and normalize
\be\label{SF6}
Z={\rm tr}~b=\sum_\tau \tilde q_\tau(\ti)=1.
\ee

The unit transfer matrix leads to a $t$-independent density matrix \eqref{N23}
\be\label{SF7}
\bar \rho\tr \T=b\tr ~,~\rho'\tr \T=b\tr.
\ee
The expectation values of local observables \eqref{N23}, \eqref{30C} are given by 
\be\label{SF8}
\kl A\T\kr =\sum_\tau A_\tau\T\rho'_{\tau\tau}\T=\sum_\tau A_\tau\T\tilde q_\tau(\ti).
\ee
If we use the same observable for all $t$, e.g. $A_\tau\T$ independent of $t$, the expectation value is independent of $t$. 

We can interpret $\rho'_{\tau\tau}\T=\tilde q_\tau(\ti)$ as the probability to find a given configuration $\tau$ at $t$. For all $t$ this probability remains the same as for $\ti$, in particular at $t_f$. For a fixed configuration $\bar\tau$ at $\ti$, e.g. $\tilde q_{\bar \tau}(\ti)=1$, $\tilde q_{\tau\neq \bar \tau}(\ti)=0$, the probability to find this configuration at $t_f$ equals unity. Thus information is transmitted completely from $\ti$ to $t_f$. 

The representations of the bilinear Grassmann operators for occupation numbers are particularly simple for the trivial model. From eq. \eqref{GB3} we infer for odd $t$
\be\label{SF9}
\tilde \B^{(\alpha)}\T=\frac{\partial}{\partial \psi_\alpha}\psi_\alpha\exp 
\left(\sum_\gamma\varphi_\gamma\psi_\gamma\right)
\ee
where $\psi=\psi\T,~\varphi=\psi\tee$. With eq. \eqref{B9a} this replaces the factor $(1+\varphi_\alpha\psi_\alpha)$ in $\tilde \K$ by $1$. We can therefore write
\be\label{SF10}
\tilde \B^{(\alpha)}\T=(1-\varphi_\alpha\psi_\alpha)\tilde \K\T. 
\ee
The same holds for the occupation numbers at $t+\epsilon$,
\be\label{SF10a}
\tilde \B^{(\bar\alpha)}\T=(1-\varphi_\alpha\psi_\alpha)\tilde \K\T. 
\ee

We infer simple expressions for the expectation values
\be\label{SF11}
\kl n_\alpha\T\kr = \int \cD\psi \cN'_\alpha\T\tilde w[\psi],
\ee
with 
\be\label{SF12}
\cN'_\alpha\T=\cN'_\alpha\tee =\bl 1-\psi_\alpha\tee \psi_\alpha\T\br.
\ee
Note that $\cN'_\alpha$ does no longer contain a factor $\K$, in contrast to $\tilde \cN'_\alpha$. In contrast to the operator $\tilde n_\alpha$ it contains no Grassmann derivatives. Since it involves an even number of Grassmann variables the place of its insertion no longer matters. 

We can also compute the Grassmann bilinear operator for the occupation number at $t$ from $\tilde \B^{c(\bar\alpha)}(t-\epsilon)$,
\ba\label{SF13}
\tilde \B^{c(\bar\alpha)}&=&\psi_\alpha\frac{\partial}{\partial \psi_\alpha}\exp 
\left(\sum_\gamma\psi_\gamma\varphi'_\gamma\right)\nn\\
&=&\psi_\alpha\varphi'_\alpha\exp \left(\sum_\gamma\psi_\gamma\varphi'_\gamma\right),
\ea
with $\varphi'_\gamma=\psi_\gamma(t-\epsilon)$. We can therefore equivalently compute $\kl n_\alpha\T\kr$ by replacing in eq. \eqref{SF11} $\cN'_\alpha$ by $\cN_\alpha$,
\be\label{SF14}
\cN_\alpha\T=\psi_\alpha\T\psi_\alpha (t-\epsilon),~\cN_\alpha\tee =\psi_\alpha(t+2\epsilon)\psi_\alpha\tee. 
\ee
One verifies that insertion of $\cN'_\alpha \T$ or $\cN_\alpha\T$ in the functional integral indeed yields the same result. The fact that $\cN'_\alpha\tee$ and $\cN'_\alpha\T$ are represented by the same observable is a particularity of the unit transfer matrix and does not hold for general $\bar S$. 

These conditions generalize to products of occupation numbers. Their expectation values obtain in the trivial model by insertions of appropriate products of $\cN_\alpha\T=\psi_\alpha\tee \psi_\alpha\T$ in the Grassmann functional integral. The insertion of the bilinears $\cN_\alpha\T$ for the computation of the expectation values of occupation numbers $n_\alpha\T$ is very convenient since this bilinear commutes with all Grassmann variables. It holds, however, only for the trivial model. Nevertheless, if the transfer matrix $\bar S$ deviates from unity only by terms $\sim\epsilon$, the errors in expectation values computed by insertion of factors $\cN_\alpha\T$ are also of the order $\epsilon$. They vanish in the continuum limit $\epsilon\to 0$. 

We may formally define a $t$-derivative
\be\label{SF15}
\psi_\gamma(t+\epsilon)-\psi_\gamma(t-\epsilon) =2\epsilon\frac{\partial}{\partial t}\psi_\gamma\T=2\epsilon\partial_t\psi_\gamma\T. 
\ee
The weight function for the trivial model is then given by 
\be\label{SF16}
\tilde w[\psi]=\exp \bl-S_E[\psi]\br\tilde b,
\ee
where 
\be\label{SF17}
S_E[\psi]=-\sum_{t}\psi_\gamma(t+\epsilon)\psi_\gamma\T=\int dt~\psi_\gamma\T\partial_t\psi_\gamma\T. 
\ee
Here we employ $\epsilon\sum_t=\int dt$. The action $S_E$ contains the typical ``time-derivative'' that appears in relativistic or non-relativistic quantum field theories for fermions. 

\subsection{Bilinear fermionic action}

Writing the weight distribution $\tilde w[\psi]$ in the form \eqref{SF16} we consider a quasi-local form of the euclidean action $S_E$,
\be\label{F13}
S_E=\sum^{t_f-\epsilon}_{t=t_{in}}{\cal L}(t),
\ee
where the Lagrangian ${\cal L}(t)$ involves only Grassmann variables at two neighboring time points $t$ and $t+\epsilon$. We take ${\cal L}(t)$ as a bilinear in the Grassmann variables, 
\be\label{F14}
{\cal L}(t)=\psi_\gamma(t)L_{\gamma\delta}\psi_\delta(t+\epsilon)=\psi_\gamma(t)B_\gamma(t+\epsilon),
\ee
with
\be\label{F15}
B_\gamma(t+\epsilon)=L_{\gamma\delta}\psi_\delta(t+\epsilon).
\ee
The partition function $Z$ is defined as a Grassmann functional integral
\be\label{F16}
Z=\int{\cal D}\psi~\tilde b ~e^{-S_E}. 
\ee

In order to find the associated model for Ising spins we need to compute the transfer matrix corresponding to 
\be\label{116A}
\tilde \K\T=\tilde \K^c\T=\exp \big\{-\cL\T\big\}. 
\ee
For odd $t$ the transfer matrix can be expressed by a Grassmann integral over $\tilde \K$,
\be\label{JU1}
\bar S\tr\T=\eta_M \int d\psi \tee d\psi\T\bar g_\tau\tee \tilde \K\T g^c_\rho\T.
\ee
We may use the identities \eqref{B19a}, \eqref{B20a},
\be\label{JU2}
\eta_M \int d\psi\T\exp \big\{ B_\gamma\tee \psi_\gamma\T\big\}g^c_\rho\T=g_\rho\bl B\tee\br,
\ee
where $g_\tau(B)$ obtains from $g_\tau(\psi)$ by replacing $\psi_\gamma\to B_\gamma$. One infers
\be\label{JU3}
\bar S\tr \T=\int d\psi \tee\bar g_\tau\bl \psi\tee \br g_\rho\bl B \tee \br.
\ee
We can expand $g_\rho(B)$ in terms of $g_\sigma(\psi)$,
\be\label{127A}
g_\rho(B)=g_\sigma(\psi)S'_{\sigma\rho}
\ee
With eq. \eqref{B25a} this yields 
\be\label{127B}
\bar S\tr \T=S'\tr.
\ee
For odd $t$ the elements of the transfer matrix correspond to the expansion coefficients of $g_\rho(B)$. 

Similarly, one has for even $t$
\be\label{JU4}
\bar S\tr\T=\eta_M \int d\psi\tee d\psi\T g'_\tau\tee \tilde \K^c\T g_\rho\T.
\ee
With 
\be\label{JU5}
\int d\psi\T \exp \big\{ B_\gamma\tee \psi_\gamma\T\big\} g_\rho\T=g^c_\rho\bl B\tee\br
\ee
this yields
\be\label{JU6}
\bar S\tr\T=\eta_M \int d\psi\tee g'_\tau\bl \psi\tee \br g^c_\rho\bl B\tee\br.
\ee
In analogy to eqs. \eqref{127A}, \eqref{127B} the elements of the transfer matrix correspond to the expansion coefficients of $g^c_\rho(B)$ in terms of $g^c_\sigma(\psi)$,
\be\label{130A}
g^c_\rho(B)=g^c_\sigma(\psi)\bar S_{\sigma\rho}. 
\ee

\subsection{Unique jump chains}

Consider now a particular form 
\be\label{JU7}
B_\gamma\tee = \delta_\gamma\psi_{\gamma'}\tee =\delta_\gamma \psi_{F(\gamma)}\tee, 
\ee
with $\gamma\to \gamma'=F(\gamma)$ an invertible map and $\delta_\gamma=\pm 1$. Up to the sign $\delta_\gamma$ the element $B_\gamma$ corresponds to one of the Grassmann variables $\psi_{\gamma'}$. In this case the $M\times M$ matrix $L$ in eq. \eqref{F15} contains in each row and each column one element $1$ or $-1$. It is a rotation matrix if $\det L=1$. The Grassmann variables $\psi_{F(\gamma)}$ and $\psi_\gamma$ are related by the linear relation (no sum over $\gamma$)
\ba\label{123A}
\psi_{F(\gamma)}=\psi_{\gamma'}=\delta_\gamma B_\gamma=\sum_\delta\delta_\gamma L_{\gamma\delta}\psi_\delta,
\ea
such that $\delta_\gamma L_{\gamma\delta}=1$ or $\delta_\gamma =$ sign $(L_{\gamma\delta})$. 

The map of Grassmann variables induces a map of the Grassmann elements
\be\label{JU8}
g_\rho(B)=\zeta_\rho g_{f(\rho)}(\psi),
\ee
with $\rho\to \rho'=f(\rho)$ again an invertible map and $\zeta_\rho=\pm 1$. The Grassmann element $g_\rho(B)$ obtains from $g_\rho(\psi)$ by replacing all $\psi_\gamma$ factors by $\delta_\gamma\psi_{\gamma'}$. The result is again one of the basis elements up to a sign. If all $\delta_\gamma$ equal one the sign $\zeta_\rho$ arises from reordering of $\psi$-factors. Otherwise, it also involves the signs $\delta_\gamma$. For the fermion model specified by eq. \eqref{JU7} the transfer matrix becomes for odd $t$
\be\label{JU9}
\bar S\tr \T=\zeta_\rho\delta_{\tau,f(\rho)}.
\ee
For even $t$ we employ that $g^c_\rho(B)$ is again a basis element of the Grassmann algebra up to a sign. If $\psi_\gamma\to\delta_\gamma \psi_{\gamma'}$ maps $g_\rho\to \zeta_\rho g_{\rho'}$, it also maps $g^c_\rho\to \zeta^c_\rho g^c_{\rho'}$. Thus the map $\rho\to f(\rho)$ is the same for $g$ and $g^c$, but the signs $\zeta_\rho$ and $\zeta^c_\rho$ may differ. For even $t$ one therefore has
\be\label{JU10}
\bar S\tr \T=\zeta ^c_\rho\delta_{\tau,f(\rho)}=\bar {\zeta}_\tau \delta_{\tau,f(\rho)}.
\ee

The transfer matrices \eqref{JU9}, \eqref{JU10} describe a unique jump chain. Every state $\rho$ at $t$ jumps to a state $f(\rho)$ at $t+\epsilon$. The transfer matrix $\bar S$ contains in each column and each line exactly one element $1$ or $-1$. This explains the second identity \eqref{JU9} where the minus signs are associated to lines rather than columns. The transfer matrix is a rotation matrix if $\det \bar S=1$. 

\subsection{Gauge freedom of signs}

The negative elements of transfer matrix for negative $\zeta_\rho$ of $\bar\zeta_\tau$ seem to be disturbing. Probability distributions for Ising spins obey eq. \eqref{N27}, such that all elements obey $\bar S\tr\geq 0$. At first sight transfer matrices \eqref{JU9}, \eqref{JU10} with negative signs $\zeta_\rho$ or $\bar\zeta_\tau$ seem not to result in positive $p[n]$. 

We observe, however, that the weight function \eqref{N14}, \eqref{N15} only depends on products of elements of the transfer matrices, and therefore not necessarily on the separate signs of elements of the individual transfer matrices. We may transform the transfer matrices by multiplication with diagonal sign matrices
\be\label{N1a}
\bar S'\T=D\tee \bar S\T D\T,~D^2\T=1,
\ee
where
\be\label{N2a}
D\tr\T=d_\tau\T\delta\tr,~d_\tau\T=\pm 1.
\ee
Provided we also transform the boundary matrix
\be\label{N3a}
b'=D(\ti)b D(t_f),
\ee
the expression \eqref{N15} and therefore the weight function \eqref{N14} remain unchanged. All transfer matrices related by sign transformations \eqref{N1a} correspond to the same weight function. The sign transformations \eqref{N1a} can be seen as local discrete gauge transformations that leave the weight distribution invariant. In particular, for a positive probability distribution $p[n]$ it is sufficient that sign matrices $D\T$ exist such that all elements of $\bar S'\T=D\tee \bar S\T D\T$ are positive. We may then employ $\bar S'\T$ for the definition of the equivalent model in the occupation number basis. 

Invariance of expectation values under gauge transformations requires that also the local bilinear operators are transformed by employing in eq. \eqref{N21} transformed matrices $B'$, 
\be\label{N4a}
B'\T=D\tee B\T D\T.
\ee
For local observables \eqref{N23} this is realized if $A_\rho\T$ or the corresponding matrix expression \eqref{30C} are kept invariant. This extends to arbitrary products. All observables involving products of occupation numbers are gauge invariant. Independently of the choice of $\bar S$ or $\bar S'$ such observables are represented by the same matrices $A'\tr$. Indeed, under the sign transformation \eqref{N1a} the restricted density matrix $\rho'$ in eq. \eqref{30A} transforms as $\rho'\T\to D\T\rho'\T D\T$. For fixed diagonal matrices $A'\T$ the trace \eqref{30B} is invariant.

The local discrete gauge symmetry \eqref{N1a} of the weight distribution is anomaly free. The diagonal sign matrices $D$ or the signs $d_\tau$ drop out in eq. \eqref{N15}. We have employed a discrete formulation which regularizes the ``functional integral'' explicitly. As long as the number of degrees of freedom remains finite the invariance of $w[n]$ is manifest. A possible continuum limit has to be taken in a way such that this property is preserved. 

\subsection{Probability distribution for unique jump chains}

For the unique jump chains \eqref{JU9}, \eqref{JU10} one can always find matrices $D\T$ such that $\bar S'\T$, as defined by eq. \eqref{N1a}, is positive. Such unique jump chains therefore correspond to generalized Ising models with a positive classical statistical probability distribution. 

In order to show this we write for $t$ odd
\be\label{N6a}
\bar S\T=\bar S' E,
\ee
with positive $\bar S'$ given by
\be\label{N7a}
\bar S'\tr=\delta_{\tau,f(\rho)}
\ee
and 
\be\label{N8a}
E_{\rho\sigma}=\zeta_\rho\delta_{\rho\sigma}.
\ee
Similarly, one uses for $t+\epsilon$ even 
\be\label{N9a}
\bar S\tee =\bar E\bar S'
\ee
with 
\be\label{N10a}
\bar E\tr =\bar\zeta_\tau\delta\tr.
\ee
For the particular case $\bar E=E$ we can realize the positive transfer matrix $\bar S'$ for all $t$ by choosing in eq. \eqref{N1a} for odd $t$ the sign matrix according to $D\T=E$, while for even $t+\epsilon$ one takes $D\tee =1$. 

For general $E$ and $\bar E$ the sequence of matrices $D\T$ can be constructed iteratively. Let us start at $\ti$. The transfer matrix $\bar S'(\ti)$ can be made positive by the choice $D(\ti)=1$, $D(\ti+\epsilon)=\bar E$. This transformation also transforms $\bar S'(\ti+\epsilon)$,
\be\label{N11a}
\bar S(\ti+\epsilon)\to \bar S(\ti+\epsilon)\bar E=\bar S' E\bar E=D(\ti+2\epsilon)\bar S'.
\ee
In the last equality we use that $\bar S'E\bar E$ has in each row and column only one element different from zero, which takes values $\pm 1$. At this stage one has
\ba\label{N12a}
&& D(\ti+\epsilon)\bar S(\ti) D(\ti)=\bar S',\nn\\
&& D(\ti+2\epsilon)\bar S(\ti+\epsilon)D(\ti+\epsilon)=\bar S',
\ea
while $\bar S(\ti+2\epsilon)$ is transformed to 
\ba\label{N13a}
&&\bar S(\ti+2\epsilon)\to \bar S(\ti+2\epsilon)D(\ti+2\epsilon)\nn\\
&&\quad =\bar E\bar S' D(\ti+2\epsilon)=D(\ti+3\epsilon)\bar S'.
\ea
With 
\be\label{N14a}
D(\ti+3\epsilon)\bar S(\ti+2\epsilon)D(\ti+2\epsilon)=\bar S'
\ee
we can then proceed iteratively and finally compute $D(t_f)$. The basic relation for this construction is the determination of $D\tee$ by 
\be\label{15a}
\bar S\T D\T=D\tee\bar S'. 
\ee
The boundary term \eqref{M8a} has to be modified to 
\be\label{N16a}
\tilde b=\eta^{\frac G2}_M g_\tau(\ti)\hat b\tr g'_\rho(t_f),
\ee
such that
\be\label{N17a}
\hat b_{\tau\sigma} D(t_f)_{\sigma\rho}=b\tr,
\ee
with $b\tr$ a positive boundary matrix. We conclude that fermion models realizing unit jump chains \eqref{F14}, \eqref{123A} can be represented in the occupation number basis by positive transfer matrices $\bar S'\tr \T=\delta_{\tau,f(\rho)}$. 

So far we have not yet specified the signs $\tilde s_\tau$ in the definition \eqref{B2a} of the basis functions $g_\tau$ of the Grassmann algebra. We have assumed so far that these signs are the same for all $t$. We may extend our discussion by choosing different signs at different $t$. A transformation $\tilde s_\tau\to \tilde s'_\tau\T$ will not affect expectation values of observables. For given $\tilde \K\T,\tilde \K^c\T$ it changes, however, the signs of $\bar S\tr\T$ in eqs. \eqref{JU1}, \eqref{JU4}. This realizes transformations of the type \eqref{N1a}. In other words there exists a choice of signs $\hat s'_\tau\T$ for which $\bar S\T=\bar S'(t)$. 

We could also follow the map in the opposite direction, starting in the occupation number basis with $\bar S'\tr=\delta_{\tau,f(\rho)}$. The corresponding factors $\tilde \K\T$ and $\tilde \K^c\T$ in the Grassmann basis are then defined by eqs. \eqref{B21a}, \eqref{B22a}. For odd $t$ one has
\ba\label{NM}
\tilde \K\T&=&\sum_\rho g_{f(\rho)}\tee g'_\rho\T,\nn\\
\tilde \K^c\tee &=&\sum_\rho g^c_{f(\rho)}(t+2\epsilon)\bar g_\rho\tee.
\ea
By suitable choices of $\tilde s_\tau\T$ one can realize eq. \eqref{116A}. 

The realization of fermionic models of the type \eqref{F14}, \eqref{123A} by Ising spins has a rather simple structure. We need to realize the transfer matrices $\bar S'$. Consider local factors $\K\T$ of the form
\be\label{M1-N}
\K\T=\exp \big\{\beta L'_{\gamma\delta}C_{\delta\gamma}\T\big\},
\ee
with 
\be\label{M2-N}
L'_{\gamma\delta}=\delta_{\delta,F(\gamma)}
\ee
related to $L_{\gamma\delta}$ in eq. \eqref{F14} by signs, $L'_{\gamma\delta}=\delta_\gamma L_{\gamma\delta}$. The factor 
\be\label{M3-N}
C_{\delta\gamma}\T=n_\delta\tee n_\gamma\T+\bar n_\delta\tee \bar n_\gamma \T-1,
\ee
with 
\be\label{M4-N1}
\bar n_\gamma\T=1-n_\gamma\T,
\ee
equals zero for $n_\delta\tee =n_\gamma\T$, either both one or both zero. For $n_\delta\tee\neq n_\gamma\T$ one finds $C_{\delta\gamma}=-1$. In the limit $\beta\to\infty$ the local factor $\K\T$ equals one if for any sequence $\rho$ of occupation numbers at $t$ the sequence of occupation numbers at $t+\epsilon$ is given by $f(\rho)$. Otherwise, one has $\K\T=0$. Here the sequence $f(\rho)$ obtains from $\rho$ by the map $\gamma\to F(\gamma)$. More precisely, denoting $\rho=\{\rho_\gamma\}=\{\rho_1\dots \rho_M\},~\rho_\gamma=0,1$, one has $f(\rho)=\{\rho_{F(\gamma)} \}=\{\rho_{F_1(\gamma)}\dots \rho_{F_M(\gamma)}\}$.  The transfer matrix obeys therefore eq. \eqref{N7a}, establishing the equivalence with the corresponding fermion model. The matrix $C$ can easily be written in terms of Ising spins $s_\gamma=2n_\gamma-1$,
\ba\label{M4-N}
C_{\delta\gamma}\T&=&-\big[n_\delta\tee \bar n_\gamma\T+\bar n_\delta \tee n_\gamma\T\big]\nn\\
&=&\frac12 \bl s_\delta \tee s_\gamma\T-1\br.
\ea
In the limit $\beta\to \infty$ the unique jump chain \eqref{M1-N} can be viewed as a particular type of cellular automata \cite{IMCA,IC,IMTH,IMTH2,IMEL}

\section{Two-dimensional fermions}
\label{Two-dimensional fermions}

Free massless two-dimensional fermions can be described by unique jump chains.  They can therefore be mapped to generalized Ising models \eqref{N27} with a positive classical statistical probability distribution. Our simple models will shed light on the emergence of a complex structure, as well as on the Lorentz symmetry characteristic for relativistic fermions. 

\subsection{Fermions and generalized Ising models in two dimensions}

Let us associate the label $\gamma$ with the position $x$ of an additional space dimension
\be\label{M5-N}
\psi_\gamma\T\to \psi(t,x). 
\ee
Occupation numbers $n(t,x)$ depend now on the positions $(t,x)$ in a two-dimensional space. For simplicity we will take $x$ periodic and choose the same distance $\epsilon$ between neighboring $x$-positions and $t$-positions, $x_0=x_{in},x_{m'}=x_{in}+m'\epsilon,x_{H}=x_{in}+H\epsilon=x_0$, with $H$ the total number of $x$-points. We consider a fermion model with 
\be\label{M6-N}
\cL\T=\sum_x\psi(t,x)\psi(t+\epsilon,x+\epsilon).
\ee
This corresponds in eq. \eqref{F14} to 
\be\label{M7-N}
L_{\gamma\delta}\to L(x,y)=\delta(x+\epsilon,y)
\ee
or 
\be\label{M8-N}
F(x)=x+\epsilon.
\ee
Moving from $t$ to $t+\epsilon$ all occupation numbers are shifted one place to the right, $x\to x+\epsilon$. This model realizes a unique jump chain. 

According to eqs. \eqref{M1-N}, \eqref{M4-N} this fermion model becomes in the occupation number basis an asymmetric Ising model with diagonal interactions only in one direction,
\be\label{R1}
\cL\T=-\frac\beta2\sum_x
\big\{s(t+\epsilon,x+\epsilon)s(t,x)-1\big\},
\ee
with $\beta\to\infty$. This simple form of the local factor \eqref{N26} has been discussed in detail in ref. \cite{CWIT}. It was argued that eq. \eqref{R1} describes the propagation of an arbitrary number of Weyl fermions in two-dimensional Minkowski space. Our map to the Grassmann formulation \eqref{M6-N} can be used to study the properties of the model in a familiar fermionic language and to reveal the presence of Lorentz symmetry. 

\subsection{Weyl fermions and complex structure}

The two-dimensional lattice of points 
\be\label{R2}
(t,x)=(\ti+n'\epsilon,x_{in}+m'\epsilon)
\ee
can be split into two sublattices. The even sublattice consists of points with $n'+m'$ even, while the odd sublattice comprises the points $n'+m'$ odd. The jumps along the diagonal remain within a given sublattice. We employ different labels for the Grassmann variables on the different sublattices, $\psi_R$ for the even sublattice, and $\psi_I$ for the odd sublattice. More precisely, we define 
\be\label{164}
\psi \x=\left\{
\begin{array}{lll}
\psi_R\x&{\rm for}&\x{\rm ~even}\\
\psi_I(t,x-\epsilon)&{\rm for}&\x{\rm ~odd}
\end{array}.
\right.
\ee
We can then write
\ba\label{R5}
&&\sum_t\sum_x\psi(t,x)\psi(t+\epsilon,x+\epsilon)=\\
&&\qquad\sum\nolimits'_{t,x}
\big[\psi_R(t,x)\psi_R(t+\epsilon,x+\epsilon)\nn\\
&&\qquad\qquad+\psi_I(t,x)\psi_I(t+\epsilon,x+\epsilon)\big].\nn
\ea
Here $\sum'_{t,x}$ sums over even lattice points. 
Only even lattice points are considered in the second line, while the number of Grassmann variables at a given point $x$ is doubled. 

A complex structure is introduced by use of a complex Grassmann variable (for even lattice points)
\be\label{R6}
\psi_C(t,x)=\psi_R(t,x)+i\psi_I(t,x).
\ee
This amounts to a map $(\psi_R,\psi_I)\to \psi_C$. Complex conjugation is a map that acts on $(\psi_R,\psi_I)$ by reversing the sign of $\psi_I$,
\be\label{R7}
K{\psi_R\choose \psi_I}=\tau_3
{\psi_R\choose \psi_I},~K\psi_C=\psi^*_C.
\ee
We can also employ a map $I$,
\be\label{R8}
I{\psi_R\choose \psi_I}={-\psi_I\choose~ \psi_R}=-i\tau_2
{\psi_R\choose \psi_I}. 
\ee
In the complex formulation it amounts to multiplication with $i$,
\be\label{R9}
I\psi_C=i\psi_C.
\ee
The pair of maps $K$ and $I$ induces a complex structure, with 
\be\label{R10}
K^2=1,~I^2=-1,~\{K,I\}=0.
\ee
The complex variables $\psi_C$ $\x$ describe a two-dimensional Weyl spinor. For a two-dimensional Majorana-Weyl spinor \cite{IMCWMW} one only retains the even sublattice in the original formulation, which amounts to $\psi_I\equiv 0$. 

As a next step we define lattice derivatives by 
\ba\label{R11}
\partial_t\psi \x&=&\frac{1}{4\epsilon}
\big\{\psi(t+\epsilon,x+\epsilon)+\psi(t+\epsilon,x-\epsilon)\nn\\
&&-\psi (t-\epsilon,x+\epsilon)-\psi(t-\epsilon,x-\epsilon)\big\},\nn\\
\partial_x\psi\x&=&\frac{1}{4\epsilon}\big\{\psi(t+\epsilon,x+\epsilon)+\psi(t-\epsilon,x+\epsilon)\nn\\
&&-\psi (t+\epsilon,x-\epsilon)-\psi(t-\epsilon,x-\epsilon)\big\}.
\ea
This yields the euclidean action $S_E$ for the complex Grassmann variables
\ba\label{R12}
S_E&=&\frac{\epsilon}{2}\sum\nolimits_{t,x}^\prime\big\{\psi^*_C\x(\partial_t+\partial_x)\psi_C\x\nn\\
&&\qquad ~~+\psi_C\x(\partial_t+\partial_x)\psi^*_C\x\big\}.
\ea
(We omit modifications of the lattice derivatives at $\ti$ and $t_f$.) 

We further may adopt a continuum notation for the sums 
\be\label{R13}
\int_{t,x}=2\epsilon^2\sum\nolimits'_{t,x}
\ee
and choose a different normalization for the complex Grassmann variables
\be\label{R14}
\psi\x=\frac{1}{\sqrt{2\epsilon}}\psi_C\x.
\ee
This yields the ``continuum expression''
\be\label{R15}
S_E=\frac12\int_{t,x}\psi^*(\partial_t+\partial_x)\psi+c.c.,
\ee
which is a familiar expression for the action of free Weyl fermions in two-dimensional Minkowski space. 

\subsection{Dirac fermions and Lorentz symmetry}

For the formulation of Dirac fermions we add another species of fermions which are now left movers. We denote by $\psi_+$ the right movers with $\cL\T$ given by eq. \eqref{M6-N}. Left movers are denoted by $\psi_-$, and the action for free Dirac fermions becomes
\ba\label{R16}
S_E&=&\sum_t\sum_x
\big\{\psi_+\x\psi_+(t+\epsilon,x+\epsilon)\nn\\
&&+\psi_-\x\psi_-(t+\epsilon,x-\epsilon)\big\}.
\ea
The equivalent asymmetric Ising model \cite{CWIT} has two types of Ising spins $s_+\x$ and $s_-\x$, with $(\beta\to\infty)$ 
\ba\label{R17}
S_E&=&-\frac\beta2 \sum_{t,x}\big\{s_+(t+\epsilon,x+\epsilon)s_+\x\nn\\
&&+s_-(t+\epsilon,x-\epsilon)s_-\x-2\big\}.
\ea
Both species have diagonal interactions, but in different directions. 

For the complex structure we assign to the points of the even sublattice $L_+$ and odd sublattice $L_-$:
\ba\label{178A}
\begin{array}{ll}
\psi_+\x=\psi_{+R}\x&{\rm for~}\x\in L_+\\
\psi_-(t,x+\epsilon)=\psi_{-R}\x&{\rm for~}(t,x+\epsilon)\in L_+\\
\psi_+(t,x+\epsilon)=\psi_{+I}\x&{\rm for ~}(t,x+\epsilon)\in L_-\\
\psi_-\x=-\psi_{-I}\x&{\rm for ~}\x\in L_-
\end{array}.
\ea
While for $\psi_+$ the sum $\sum'_{t,x}$ extends over the points of the even sublattice, we take for $\psi_-$ only the points of the odd sublattice. We employ again the lattice derivatives \eqref{R11} and the normalization \eqref{R14}, such that eq. \eqref{R15} is extended to 
\be\label{R18}
S_E=\frac12\int_{t,x}\big\{\psi^*_+(\partial_t+\partial_x)\psi_++\psi^*_-
(\partial_t-\partial_x)\psi_-+c.c.\big\}.
\ee
In the following we simplify the notation to 
\be\label{R19}
S'_E=\int_{t,x}\big\{\psi^*_+(\partial_t+\partial_x)\psi_++\psi^*_-
(\partial_t-\partial_x)\psi_-\big\},
\ee
where it is understood that $S_E=(S'_E+S^{\prime *}_E)/2$. (If partial integration in a continuum formulation is allowed one has $S_E=S'_E$.)

In order to see the Lorentz-symmetry we introduce a two-component complex spinor
\be\label{R20}
\psi={\psi_+\choose \psi_-}.
\ee
The action \eqref{R19} can be written in terms of Dirac matrices
\ba\label{R21}
S'_E&=&\int_{t,x}\psi^\dagger(\partial_t+\tau_3\partial_x)\psi=-\int_{t,x}\bar\psi\gamma^\mu\partial_\mu\psi\nn\\
&=&-\int_{t,x}\bar\psi(\gamma^0\partial_t+\gamma^1\partial_x)\psi. 
\ea
Here we express the two-dimensional Dirac matrices in terms of Pauli matrices
\be\label{R22}
\gamma^0=-i\tau_2~,~\gamma^1=\tau_1,
\ee
with 
\be\label{R23}
\{\gamma^\mu,\gamma^\nu\}=2\eta^{\mu\nu}
\ee
and Minkowski metric $\eta^{\mu\nu}=$diag$(-1,1)$. We also define
\be\label{R24}
\bar \psi=\psi^\dagger \gamma^0=(\psi^*_-,-\psi^*_+).
\ee
The analogue of $\gamma^5$ in four dimensions is 
\be\label{R25}
\bar\gamma=-\gamma^0\gamma^1=\tau_3,~\{\bar\gamma,\gamma^\mu\}=0,
\ee
such that the right movers (with positive momentum) obey $\bar\gamma\psi =\psi$, while the left movers correspond to the negative eigenvalue of $\bar\gamma,\bar\gamma\psi=-\psi$. (The sign of $\gamma^0$ differs from the conventions of ref. \cite{IMCWF}.) The Minkowski action $S_M$ in eq. \eqref{IN1} is related to $S'_E$ by 
\be\label{182A}
S_M=-iS'_E.
\ee

Infinitesimal Lorentz transformations act on $\psi$ as
\be\label{R26}
\delta\psi=-\eta\Sigma^{01}\psi,~\delta\bar\psi=\eta \bar\psi \Sigma^{01},
\ee
with generator 
\be\label{R27}
\Sigma^{01}==\frac14[\gamma^0,\gamma^1]=\frac12\tau_3.
\ee
The two Weyl spinors transform separately, by opposite multiplicative scalings, 
\be\label{R28}
\delta\psi_+=-\frac\eta2\psi_+,~\delta\psi_-=\frac{\eta}{2}\psi_-.
\ee

For continuous spacetime the Lorentz transformations act also on the coordinates $t$ and $x$, $\delta_{\cL}=\delta+\delta_\xi$, with $\delta_\xi$ reflecting the transformation of coordinates. With respect to the coordinate transformations $\partial_\mu\psi$ transforms as a two-dimensional Lorentz vector, or 
\be\label{R29}
\delta_\xi\partial_t\psi=\eta\partial_x\psi,~\delta_\xi\partial_x\psi=\eta\partial_t\psi,
\ee
resulting in 
\ba\label{R30}
\delta_\xi(\partial_t+\partial_x)\psi=\eta(\partial_t+\partial_x)\psi,\nn\\
\delta_\xi(\partial_t-\partial_x)\psi=-\eta(\partial_t-\partial_x)\psi. 
\ea
It is easy to check that terms like $\psi^*_+(\partial_t+\partial_x)\psi_+$ or $\psi^*_-(\partial_t-\partial_x)\psi_-$ are invariant under the combined transformations \eqref{R28}, \eqref{R30}. We observe that $\psi_+$ and $(\partial_t+\partial_x)\psi_+$ scale oppositely with respect to the combined transformation
\be\label{R31}
\delta_{\cL}\psi_+=-\frac\eta2\psi_+,~\delta_{\cL}(\partial_t+\partial_x)\psi_+=\frac{\eta}{2}(\partial_t+\partial_x)\psi_+,
\ee
and similarly for $\psi_-$
\be\label{193A}
\delta_{\cal L} \psi_-=\frac\eta2\psi_-,~\delta_{\cal L}(\partial_t-\partial_x)\psi_-=-\frac\eta2(\partial_t-\partial_x)\psi_-.
\ee
The Lorentz invariance of the action \eqref{R19} is manifest.

The spacetime part $\delta_\xi$ of the Lorentz transformation can be realized by a change of distances between lattice points. Let us consider different lattice distances $\epsilon_t$ and $\epsilon_x$ for the $t$ and $x$ coordinates respectively, e.g.
\be\label{LD1}
\x=(\ti+n'\epsilon_t,~x_{in}+m'\epsilon_x).
\ee
For a suitable definition of lattice derivatives and a suitable change of $\epsilon_t$ and $\epsilon_x$ one can realize $\delta_\xi$ with the property \eqref{R29}. The spacetime part $\delta_\xi$ of the Lorentz transformation is a particular case of the lattice diffeomorphism transformations \cite{CWLDS}.

A type of Lorentz transformation can also be implemented for the Ising spins in the occupation number basis. Expressing the action \eqref{R17}  in terms of suitable lattice derivatives, and transforming $s_\pm$ similar to $\psi_\pm$ in eq. \eqref{R28}, one realizes Lorentz invariance of the action. The translation to Lorentz-symmetry for the transfer matrix remains to be worked out. For $\beta\to\infty$ the fermion-bit map reveals the presence of Lorentz symmetry in the generalized Ising model \eqref{R17}. This extends to other symmetries of the fermion model,  as the global $U(1)$-gauge transformation associated to conserved fermion number.

\section{Conclusions}
\label{Conclusions}

We have constructed a very general ``local'' map between Ising spins and Grassmann variables, that extends to weight functions, ``functional integrals'', observables and their expectation values. Fermion models based on Grassmann variables and generalized Ising models should be considered as two different representations of the same physics, namely the fermion basis or the occupation number basis. 

All fermion models based on local Grassmann variables can be represented by an appropriate weight function for local Ising spins. This includes fermionic quantum field theories with a Minkowski signature. It also comprises fermionic quantum field theories with euclidean signature, that arise in the analytic continuation of the ones in Minkowski space, as for lattice gauge theories. The two cases will lead to different weight functions in the occupation number basis. A systematic study of their relation will be highly interesting. As the examples of sect. \ref{Two-dimensional fermions} demonstrate, it is a priori not obvious which case leads to a positive probability distribution $p[n]$. While the Minkowski-space action \eqref{IN1} leads to a positive probability distribution \eqref{IN2}, the properties of the analytic continuation according to ref. \cite{IMCWES} still need to be established. 

The class of fermion models that lead to a positive probability distribution in the occupation number basis are of particular interest. They can be described by classical statistical systems for Ising spins, with many highly developed methods available. At the present stage only very simple examples have been discussed, while a classification of models with positive weight distribution remains open. It is rather straightforward to construct for a given fermion model the transfer matrix $\bar S$, at least formally. There are, however, large equivalence classes of transfer matrices that all lead to the same weight distribution $w[n]$. The  gauge freedom of signs discussed in sect. \ref{Simple fermion models} is only a small subset of the equivalence transformations discussed in ref. \cite{CWIT}. As a consequence, negative elements of the transfer matrix can be compatible with a positive weight function, rendering the classification problem difficult. This concerns, in particular, fermionic models that can no longer be represented as unique jump chains. The opposite way may be simpler. For any classical statistical generalized Ising model \eqref{3B} one can construct an associated fermion model with alternating evolution factors $\tilde \K\T$ and $\tilde \K^c\T$. One may then investigate in the fermionic formulation if such a model can be brought to a simple form by appropriate similarity transformations.

Already our simple explicit models demonstrate that several prejudices concerning the relation between quantum physics and classical statistics are inappropriate. \\
(i) It is often believed that fermions are typical quantum objects that cannot be formulated in classical statistics. The fermion-bit map proves this to be wrong. \\
(ii) The probabilistic information in quantum mechanics is encoded in a wave function or density matrix, while classical statistics involves probabilities. A ``classical wave function'' and associated ``classical density matrix'' arises in classical statistics as well \cite{IMCWCW}. It is a crucial quantity if information transport between different $t$-layers or hypersurfaces is studied \cite{CWIT}.\\
(iii) It is often said that quantum theory is the ``physics of phases'', with emphasis on the complex character of the wave function. A complex structure can, however, arise naturally in classical statistics, as shown in our example in sect. \ref{Two-dimensional fermions}.\\
(iv) Quantum physics involves non-commuting operators, while classical statistics is often assumed to describe only commuting products of observables. The transfer matrix, the local bilinear operators and the density matrix are all structures in classical statistics. Corresponding matrix products are often non-commuting.\\
(v) Time, the unitary evolution and the Minkowski signature in quantum field theories seem to play a particular role for quantum physics that is not visible in classical statistics. We have established the existence of simple generalized Ising models on euclidean lattices that describe the unitary time-evolution in fermionic quantum field theories in Minkowski space. 

With these observation one may ask if quantum physics is only a subset of classical statistics \cite{IMCWEQM}. General theorems \cite{IMBE,IMKS} seem to contradict this possibility. The issue of Bell's inequalities \cite{IMBE} concerns the choice of the appropriate conditional probabilities for the description of sequences of measurements, or more generally, the choice of appropriate correlation functions for measurements of two observables. The ``classical correlation function'' obeys Bell's inequalities. It is, however, not appropriate for sequences of measurements in subsystems that are independent of the environment. Appropriate correlation functions that only involve information for the subsystem violate \cite{IMCWEQM} Bell's inequalities, in accordance with experiment. The Kochen-Specker theorem \cite{IMKS} assumes that a unique classical observable can be associated to every quantum observable. This assumption is not realized for the ``incomplete statistics'' \cite{IMCWOQM,IMCWEQM} of subsystems. 

The discussion of two-dimensional fermions in sect. \ref{Two-dimensional fermions} realizes a quantum system in a classical statistical setting. It is too simple, however, in order to exhibit the characteristic interference phenomena of quantum physics. It will be highly interesting to see if quantum models with interference, as massive fermions in a potential, can find a description in terms of classical statistical generalized Ising models.

\medskip\noindent
{\em Acknowledgment:} This work is part of and supported by the DFG Collaborative Research Centre ``SFB 1225 (ISOQUANT)''


\vspace{2.0cm}\noindent

\bibliography{Fermions_as_generalized_Ising_models}

\end{document}